\documentclass[aps,pra,reprint,superscriptaddress,nofootinbib,longbibliography,floatfix]{revtex4-2}
\usepackage[T1]{fontenc}
\usepackage[utf8]{inputenc}
\usepackage{lmodern}
\usepackage{array}[=2016-10-06]
\usepackage{amsmath,amssymb,bm,graphicx,booktabs,mathtools}
\usepackage{placeins}
\usepackage{hyperref}
\ifdefined\pdfinfoomitdate\pdfinfoomitdate=1\fi
\ifdefined\pdftrailerid\pdftrailerid{}\fi
\ifdefined\pdfsuppressptexinfo\pdfsuppressptexinfo=15\fi
\hypersetup{colorlinks=true,linkcolor=blue,citecolor=blue,urlcolor=blue,pdftitle={Exceptional points in non-Markovian quantum-state diffusion: From coefficient dynamics to state signatures},pdfauthor={Yusui Chen, Hang Liu, Kaiqi Xiong, Quanzhen Ding, Eric Guo}}
\newcommand{\ii}{\mathrm{i}}
\newcommand{\dd}{\mathrm{d}}
\newcommand{\Id}{\mathbb I}
\newcommand{\sinhc}{\operatorname{sinhc}}
\newcommand{\Cov}{\operatorname{Cov}}
\newcommand{\diag}{\operatorname{diag}}
\newcommand{\Rea}{\operatorname{Re}}
\newcommand{\Ima}{\operatorname{Im}}
\newcommand{\tr}{\operatorname{tr}}
\newcommand{\norm}[1]{\left\lVert#1\right\rVert}
\newcommand{\ket}[1]{\lvert#1\rangle}
\newcommand{\bra}[1]{\langle#1\rvert}

\begin{document}
\title{Exceptional points in non-Markovian quantum-state diffusion: From coefficient dynamics to state signatures}
\author{Yusui Chen}
\email{yusui.chen@nyit.edu}
\affiliation{Department of Physics, College of Arts and Sciences, New York Institute of Technology, Old Westbury, NY 11568, USA}
\author{Hang Liu}
\affiliation{Department of Electrical and Computer Engineering, Rutgers University, Piscataway, NJ 08854, USA}
\author{Kaiqi Xiong}
\affiliation{Intelligent Cyber \& Network Security (ICNS) Lab and Cyber Florida, University of South Florida, Tampa, FL 33620, USA}
\author{Quanzhen Ding}
\affiliation{Department of Physics, Stevens Institute of Technology, Hoboken, NJ 07030, USA}
\author{Eric Guo}
\affiliation{High Technology High School, Lincroft, NJ 07738, USA}
\date{November 8, 2026}
\begin{abstract}
An exceptional point of a memory-coefficient equation need not have the same signature in every quantum-state observable. We examine this relation within exact non-Markovian quantum-state diffusion (NMQSD), following the same operator closure from its coefficient dynamics to the non-Markovian master equation obtained by noise averaging. A finite homogeneous lift identifies the Jordan chains reached by the causal initialization, while state reconstruction determines which contributions survive in a specified readout. Two models expose the distinction. For common-bath qubits restricted to zero and one excitation, a noise-independent lowering-form $O$ operator gives a closed master equation. Every defective decaying mode transfers its full pole order to a suitable unconditional excitation--reference coherence. A known third-order configuration illustrates both a quadratic--exponential coherence and cancellation of its highest-order term in another quadrature. For a position-coupled oscillator in a fixed-photon-number optomechanical sector, the exact closure contains a scalar noise-history term. On the regular normalizable branch studied for bath decay rates below the oscillator frequency, two second-order coefficient blocks lead to a double pole in the conditional displacement response to the NMQSD noise history, at long preparation age and fixed lag. The purely algebraic coefficient tail cancels from the conditional Gaussian width. Complete noise averaging changes the master-equation coefficients and combines conditional width with the weighted spread of state centers. Together, the models show how causal preparation, state reconstruction, and averaging give coefficient exceptional points a precise, readout-dependent meaning.

\end{abstract}
\maketitle
\section{Introduction}
\label{sec:introduction}

A reservoir with finite correlation time can return information to an open quantum system and alter its relaxation, coherence, and correlations \cite{breuer2016,devega2017}. These effects depend on the reservoir structure as well as on the system--bath coupling. Examples include entanglement revivals for qubits coupled to independent reservoirs \cite{bellomo2007} and noise-assisted coherence protection in hierarchical environments \cite{zhaoma2022}. In a piezoelectric optomechanical model, Ding et al. reported singular coefficient functions in non-Markovian quantum-state diffusion (NMQSD) and distinct entanglement dynamics as the bath central frequency was varied \cite{ding2021}. Such coefficient behavior motivates a spectral question: can a degeneracy in the equations that encode environmental memory be identified through the quantum state that they generate?

Exceptional points (EPs), at which eigenvalues and eigenvectors of a non-Hermitian operator coalesce, provide a natural setting for this question \cite{heiss2012,ashida2020}. Their Jordan structure produces polynomial factors multiplying exponential evolution. In an open quantum system, the operator whose spectrum is studied must be specified: effective Hamiltonians and Liouvillians can have different EPs because quantum jumps affect the density-operator evolution \cite{minganti2019}. With reservoir memory, pseudomode and hierarchical descriptions explicitly project extended Jordan modes into reduced-state dynamics \cite{lin2025}, while reaction-coordinate treatments connect poles to observable-dependent polynomial factors and late-time relaxation \cite{holst2026}. These physical connections build on enlarged-space representations of structured reservoirs \cite{garraway1997,tamascelli2018,pleasance2020}. Zhang et al. describe an experimental superposition preparation and joint qubit--pseudomode tomography that resolves non-Markovian EPs with quantum jumps retained \cite{zhang2025}. Different coherence functions can display different orders \cite{arkhipov2020}, so the preparation and readout matter as well as the spectrum.

The NMQSD formulation follows these connections through the stochastic system state \cite{strunz1996,diosi1997,dgs1998,strunz1999}. In this formulation, the reservoir enters a stochastic pure-state equation through colored Gaussian noise and a functional derivative. An $O$ operator represents that derivative and, for suitable system algebras, obeys a finite exact closure. Exact time-local equations have been constructed for interacting qubits and multilevel systems \cite{q_Zhao2011,jing2012}. Averaging the same stochastic equation yields the reduced density operator; the corresponding master equation has been derived explicitly for quantum Brownian motion and multiple qubits \cite{strunz2004,chen2014}. This established route carries the coefficient spectrum through state reconstruction and noise averaging within the same reservoir model.

Here we ask which features of a coefficient EP survive the state map and complete noise average in two exactly closed models. A linear homogeneous lift makes the nonlinear coefficient dynamics accessible to spectral analysis, but its physical solution has a prescribed initial vector and a denominator projection. These select the reached Jordan chains. State reconstruction can retain, combine, or cancel their contributions. The noise average must also include correlations between the $O$ operator and the stochastic projector; a finite coefficient closure does not make this step automatic.

For common-bath qubits in the zero- and single-excitation sector, the $O$ operator is noise independent. We establish an affine-transpose relation between the coefficient and state-memory matrices and prove that every defective decaying mode retains its full pole order in a suitable unconditional excitation--reference coherence. The proof covers arbitrary Hermitian single-excitation Hamiltonians and nonzero collective rank-one coupling with a single-exponential bath correlation. A known third-order EP (EP3) configuration \cite{lin2025} illustrates the general result and cancellations in individual quadratures. A determinant identity then relates the coefficient chart to the amplitude map. On intervals where this map is invertible, its inversion identifies the memory coefficients from separately prepared ensembles and known model parameters.

The position-coupled oscillator tests the extra averaging step that the qubit closure avoids. In driven, linearized red-sideband optomechanics, reservoir memory shifts a hybrid second-order EP (EP2) that enters the probe-reflection response \cite{ghosh2026}. Here the photon number is conserved, with no drive or optical loss, and radiation pressure supplies an affine force. The exact $O$ operator contains a scalar noise-history term \cite{dgs1998}. On the regular normalizable branch studied here, the two coefficient roots produce a double-pole conditional displacement response to the NMQSD noise history at coalescence, in the long-preparation-age limit at fixed lag. The purely algebraic coefficient tail cancels from the Gaussian width. The full operator average then determines the effective master equation. Comparing the conditional and reduced-state covariances identifies how the spread of state centers changes the fluctuations after averaging. Together, these results give a readout-dependent interpretation of coefficient EPs within the exact NMQSD-to-master-equation route.

\section{Common coefficient-to-state framework}
\label{fw:section}
\label{sec:framework}
\label{fw:representation}

Using $\hbar=1$, the linear NMQSD equation for an initially factorized system and Gaussian vacuum reservoir reads \cite{dgs1998,strunz2004}
\begin{align}
\partial_t|\psi_t\rangle
 &=\left[-\mathrm iH_S+Lz_t^*-L^\dagger\bar O_t\right]|\psi_t\rangle,
 \label{fw:qsd}\\
\bar O_t&=\int_0^t\alpha(t,s)O(t,s,z^*)\,\mathrm ds,
\nonumber\\
\frac{\delta|\psi_t\rangle}{\delta z_s^*}&=O(t,s,z^*)|\psi_t\rangle.
\nonumber
\end{align}
Here $H_S$ is the system Hamiltonian and $L$ is the system operator that couples to the reservoir. The reference noise has zero mean, $\mathbb E_0[z_tz_s^*]=\alpha(t,s)$, and $\mathbb E_0[z_tz_s]=0$. Both applications use the full prescribed correlation
\begin{equation}
\alpha(t,s)=A_0e^{-\gamma|t-s|-\mathrm i\Omega(t-s)},
\qquad d=\gamma+\mathrm i\Omega,
\label{fw:kernel}
\end{equation}
with $A_0,\gamma>0$ and real $\Omega$. The parameters set the correlation amplitude, inverse memory time, and central frequency. We use this full kernel as the definition of the effective reservoir. Truncating its full-line Lorentzian transform to positive frequencies would change the prescribed correlation. The system preparations are specified separately.

In the vacuum Bargmann representation, fixed coherent-state labels generate the colored process
\begin{align}
z_t^*&=-\ii\sum_k g_kz_k^*e^{\ii\omega_kt},\qquad \mathbb E_0[z_t^*]=0,\nonumber\\
\mathbb E_0[z_tz_s^*]&=\alpha(t,s),\qquad \mathbb E_0[z_tz_s]=0.
\label{o:eq:noise}
\end{align}
Here $g_k$ and $\omega_k$ denote the mode weights and frequency labels of a Bargmann representation chosen to reproduce the prescribed correlation in Eq.~\eqref{fw:kernel}; its continuum realization is understood at the level of this full kernel. The reference Gaussian measure satisfies $\mathbb E_0[z_kz_l^*]=\delta_{kl}$. Bath annihilation operators become derivatives with respect to $z_k^*$, and the chain rule gives the history derivative in Eq.~\eqref{fw:qsd}. Thus the ket depends on the entire label configuration, not only on the instantaneous value $z_t^*$.

Commuting the history derivative with time evolution gives the consistency equation
\begin{equation}
\partial_tO=[-\ii H_S+Lz_t^*-L^\dagger\bar O_t,O]
-L^\dagger\frac{\delta\bar O_t}{\delta z_s^*},\quad O(t,t,z^*)=L.
\label{o:eq:consistency}
\end{equation}
An exponential correlation simplifies the integrated coefficient equations, but a finite exact closure also requires a suitable operator algebra. For the qubits, the retained excitation sector closes on lowering operators \cite{chen2014}. For the oscillator, closure requires a scalar noise-history term as well as position and momentum \cite{dgs1998,strunz2004}.

The exact density operator and normalized-state ensemble obey
\begin{equation}
\rho_t=\mathbb E_0[|\psi_t\rangle\langle\psi_t|],
\qquad
\mathrm d\mathbb P_t=\langle\psi_t|\psi_t\rangle\,\mathrm d\mathbb P_0,
\label{fw:average}
\end{equation}
where $\mathbb E_0\langle\psi_t|\psi_t\rangle=1$ for a normalized preparation. Thus averaging normalized kets with the reference Gaussian weight is generally insufficient. Equation~\eqref{fw:average} defines a fixed-time ensemble. For general non-Markovian dynamics, these pure-state ensembles cannot be joined into pure-state trajectories generated by continuous monitoring of the environment \cite{wiseman2008}.

Writing $\Pi_t=|\psi_t\rangle\langle\psi_t|$ and retaining the operator--state correlations gives the formally exact non-Markovian master equation (NMME) \cite{strunz2004,chen2014},
\begin{align}
\dot\rho_t&=-\mathrm i[H_S,\rho_t]
 +[L,\mathcal A_t^\dagger]+[\mathcal A_t,L^\dagger],
\label{fw:nmme}\\
\mathcal A_t&=\mathbb E_0[\bar O_t\Pi_t].
\nonumber
\end{align}
This is the NMQSD-to-master-equation relation used in both applications. If $\bar O_t$ is independent of the noise labels, $\mathcal A_t=\bar O_t\rho_t$ and the equation closes directly. For a noise-dependent $O$ operator, the joint average must instead be evaluated completely. Zero-mean noise does not imply zero noise--state correlations. In the oscillator calculation these correlations carry the scalar history sector into the effective NMME coefficients. This distinction also appears in general Gaussian stochastic and master-equation constructions \cite{diosi2014,ferialdi2016}.

\subsection{Projective coefficients and the causal lift}
\label{fw:lift}

The deterministic coefficient cores derived below have the form
\begin{equation}
\dot{\boldsymbol y}=\boldsymbol b+K\boldsymbol y
 -\boldsymbol y\bigl(\boldsymbol r^T\boldsymbol y\bigr),
\qquad \boldsymbol y(0)=0,
\label{fw:riccati}
\end{equation}
with constant, generally complex entries. The transpose denotes a bilinear coordinate pairing, not a Hermitian inner product. Introducing homogeneous coordinates gives
\begin{equation}
\begin{gathered}
\boldsymbol y=\frac{\boldsymbol x}{q},\qquad
v=\binom{\boldsymbol x}{q},\qquad
\dot v=Mv,\qquad v(0)=e_q,\\
M=\begin{pmatrix}K&\boldsymbol b\\\boldsymbol r^T&0\end{pmatrix},
\qquad
\frac{\dot q}{q}=\boldsymbol r^T\boldsymbol y.
\end{gathered}
\label{fw:homogeneous}
\end{equation}
Here $e_q$ is the last coordinate vector. Differentiating $\boldsymbol x/q$ proves Eq.~\eqref{fw:riccati} wherever $q\ne0$. An eigenvector of $M$ with nonzero last coordinate gives a finite stationary coefficient vector $\boldsymbol y_*$; its eigenvalue is $\boldsymbol r^T\boldsymbol y_*$. Eigenvectors with vanishing last coordinate lie outside this chart.

This representation also separates spectral branching from finite-time singular behavior. If the entries of $M$ depend analytically on a control parameter, the finite-time matrix exponential is analytic even at an exceptional point. Individual labeled eigenvalues can have fractional-power splittings while their combined contribution to $v(t)$ remains regular. The ratios inherit that local regularity wherever $q\ne0$. Consequently, identifying an exceptional point from a defective matrix does not require a discontinuity in the finite-time coefficient trajectory. Conversely, a pole caused by $q=0$ does not establish a spectral degeneracy.

A zero of $q$ therefore marks failure of the finite coefficient coordinates. The homogeneous solution remains defined at every finite time. Regularity of the reconstructed physical state requires the model-specific state map and is not a consequence of this coordinate observation alone. In the qubit model that map yields an exact determinant identity; in the oscillator model the Gaussian state and its averaging chart must also be controlled.

The prescribed zero-coefficient preparation fixes both the initial vector and the denominator readout. The Laplace transform $\widetilde q(s)$ of $q(t)$, with complex frequency variable $s$, is
\begin{equation}
\widetilde q(s)=e_q^T(sI-M)^{-1}e_q.
\label{fw:qresolvent}
\end{equation}
An eigenvalue degeneracy is insufficient to fix the late-time coefficient tail. Suppose that $\widetilde q$ has a pole of order $r$ at $\mu_*$, whose real part is strictly larger than that of every other pole of this scalar function. Partial fractions then give
\begin{equation}
\begin{gathered}
q(t)=e^{\mu_*t}\left[a_{r-1}t^{r-1}+O(t^{r-2})\right]
       +R_{\rm sub}(t),\\
\boldsymbol r^T\boldsymbol y(t)=\frac{\dot q(t)}{q(t)}
 =\mu_*+\frac{r-1}{t}+O(t^{-2}).
\end{gathered}
\label{fw:tail}
\end{equation}
Here $a_{r-1}\ne0$ and $R_{\rm sub}$ is exponentially subleading relative to the first term. For $r=1$, the leading polynomial is constant. The second line follows by logarithmic differentiation. The pole order must survive both excitation from $e_q$ and projection back onto $e_q$; it need not equal the largest Jordan block of $M$. Equal-real-part poles or a subleading exceptional point do not satisfy the stated dominance assumption. The two applications explicitly establish the required noncancellation for their causal preparations.

A reached dominant pole of order $r$ therefore contributes $(r-1)/t$ to the contraction $\boldsymbol r^T\boldsymbol y$. The sign and scale of a named model coefficient follow from its relation to this contraction, as derived in the corresponding model section. State reconstruction and the specified readout then determine an observable's polynomial degree, decay rate, and late-time dominance; their propagation matrix and projection can differ from those of the coefficient lift.

\subsection{A specified readout selects the Jordan contribution}
\label{fw:projection}

State reconstruction supplies the appropriate propagation operator for a proposed signal. When that particular problem reduces to a constant linear system, write
\begin{equation}
\dot u=Au+Bf,\qquad o=Cu,\qquad
\mathcal T(s)=C(sI-A)^{-1}B.
\label{fw:transfer}
\end{equation}
For zero initial response, $Ce^{At}B$ is the impulse-response kernel. The same expression propagates an initial preparation of the form $Bu_0$. The matrices $A,B,C$ must be derived for the stated input, preparation, and output; $A$ is not assumed to be the coefficient matrix $M$.

Let $P_\lambda$ be the spectral projector onto the generalized eigenspace of $A$ at $\lambda$, and put $N_\lambda=(A-\lambda I)P_\lambda$. If its nilpotency index is $r_\lambda$, the contribution of this eigenspace is
\begin{equation}
\mathcal T_\lambda(s)=
\sum_{j=0}^{r_\lambda-1}
\frac{CN_\lambda^jP_\lambda B}{(s-\lambda)^{j+1}}.
\label{fw:residues}
\end{equation}
This follows by expanding the inverse of $(s-\lambda)I-N_\lambda$ as a finite geometric series. Equivalently,
\begin{equation}
Ce^{At}P_\lambda B
=e^{\lambda t}\sum_{j=0}^{r_\lambda-1}
\frac{t^j}{j!}\,CN_\lambda^jP_\lambda B.
\label{fw:jordanreadout}
\end{equation}
The highest-order pole survives precisely when $CN_\lambda^{r_\lambda-1}P_\lambda B\ne0$. For matrix-valued inputs or outputs, this guarantees a suitable pair of channels, not every preparation and readout. A particular quadrature can cancel the highest term while retaining lower-order terms. These are standard linear-algebra identities underlying observable-dependent EP signatures \cite{lin2025,arkhipov2020}; the physical work is to identify and justify the relevant map.

For the qubits, the relevant input is a coherent system preparation and the output is an unconditional excitation--reference coherence. For the oscillator, the response variable is the stochastic displacement, and its constant propagation matrix is obtained only after a long-preparation-age limit at fixed lag. The corresponding input varies the NMQSD history coordinate. These maps will be derived from the respective state equations.
Neither Eq.~\eqref{fw:residues} nor coefficient spectral similarity replaces nonlinear state reconstruction or the average in Eq.~\eqref{fw:average}. A Gaussian width can cancel a coefficient tail, and a reduced covariance combines conditional widths with the weighted spread of state centers. The two applications below show explicitly where these operations preserve the Jordan contribution and where they change its signature. The analysis concerns exact closures for an initially factorized vacuum reservoir with a prescribed exponential correlation; the spectral statements are not a general criterion for information backflow or for an EP of the instantaneous reduced-state generator.

\section{Common-reservoir multiqubit systems}
\label{sec:qubits}

A common reservoir can couple several system transitions through one lowering operator. Restricting the system to at most one excitation makes its exact $O$ operator noise independent. This provides a first test of the coefficient-to-state connection: the same deterministic row enters both the coefficient EP and the unconditional system coherence.

\subsection{Exact NMQSD coefficients and their linear lift}
\label{q:sec:model}
Consider $N_q$ qubits restricted to their zero- and single-excitation
sector. The state $\ket{G_S}=\ket{0\cdots0}$ is the system's zero-excitation reference, distinct from the reservoir vacuum, and $\{\ket j\}_{j=1}^{N_q}$ spans its single-excitation subspace. Subtracting the zero-excitation energy from the entire system Hamiltonian gives $H_S=0\oplus H_1$, where the arbitrary time-independent Hermitian matrix $H_1$ contains transition energies relative to that reference.
The common lowering operator is
\begin{equation}
 L=\sum_{j=1}^{N_q}\kappa_j\ket{G_S}\bra j,\qquad
 \kappa=(\kappa_1,\ldots,\kappa_{N_q}).
 \label{q:eq:L}
\end{equation}
We choose $L$ and $\kappa$ dimensionless. The coefficients $\kappa_j$
weight system transitions in $L$, whereas the microscopic bath
weights $g_k$ specify the reservoir correlation in
Eq.~\eqref{fw:kernel}. The interaction conserves total excitation
number, and the initially independent reservoir is in its Gaussian
vacuum with that exponential correlation. Thus $A_0$ has units of
frequency squared and $d=\gamma+i\Omega$.

Within this invariant sector, the $O$ operator closes exactly as
\begin{equation}
 \begin{aligned}
 O(t,s)&=\sum_j f_{q,j}(t,s)\ket{G_S}\bra j,\\
 \bar O_t&=\sum_j F_{q,j}(t)\ket{G_S}\bra j,\\
 \bm F_q(t)&=\int_0^t\alpha(t,s)\bm f_q(t,s)\,ds .
 \end{aligned}
 \label{q:eq:qsd}
\end{equation}
The rows $\bm f_q$ and $\bm F_q$ are noise independent, and both
operators annihilate $\ket{G_S}$. This property uses the vacuum
preparation and excitation restriction. General multiexcitation
qubit closures contain additional noise terms
\cite{q_Zhao2011,chen2014}.
The consistency equation and $\bm f_q(t,t)=\kappa$ give
\begin{align}
 \partial_t\bm f_q&=i\bm f_qH_1+
       (\bm f_q\kappa^\dagger)\bm F_q,\label{q:eq:f}\\
 \dot{\bm F}_q&=A_0\kappa-d\bm F_q+i\bm F_qH_1+
       (\bm F_q\kappa^\dagger)\bm F_q,\nonumber\\
 &\hspace{5em}\bm F_q(0)=0 .
 \label{q:eq:riccati}
\end{align}
Appendix~\ref{q:app:prior} derives this closure from the consistency
equation. The coefficient spectrum follows directly from this exact
NMQSD initial-value problem. Set $\bm F_q=\bm x^T/q$ on $q\ne0$.
Its linear lift is
\begin{equation}
 \begin{aligned}
 \dot v&=M_qv,\qquad v(0)=e_a=\binom{0_{N_q}}{1},\\
 M_q&=
 \begin{pmatrix}
 -dI_{N_q}+iH_1^T&A_0\kappa^T\\
 -\kappa^*&0
 \end{pmatrix}.
 \end{aligned}
 \label{q:eq:lift}
\end{equation}
Here $e_a$ is the last-coordinate unit column, and
$c_q=\bm F_q\kappa^\dagger=-\dot q/q$.
A finite stationary direction with eigenvalue $\mu$ gives
\begin{equation}
 \begin{aligned}
 \bm F_{q,*}(\mu)
 &=A_0\kappa[(\mu+d)I-iH_1]^{-1},\\
 c_{q,*}&=-\mu ,
 \end{aligned}
 \label{q:eq:fixed}
\end{equation}
whenever the displayed inverse exists. Coefficient EPs are
identified from the defective eigenvalues of $M_q$.

\subsection{A third-order EP of the coefficient equations}
\label{q:sec:example}
For $N_q=2$, take $\ket{G_S}=\ket{00}$, $\ket1=\ket{10}$,
$\ket2=\ket{01}$, $A_0=\gamma^2$, $\Omega=0$, and
\begin{equation}
 \begin{aligned}
 H_1&=\gamma
 \begin{pmatrix}
 1/6&1/(6\sqrt3)\\
 1/(6\sqrt3)&-1/6
 \end{pmatrix},\\
 \kappa&=\left(\frac{\sqrt6}{9},\frac{\sqrt2}{3}\right).
 \end{aligned}
 \label{q:eq:params}
\end{equation}
The characteristic polynomial calculated from
Eq.~\eqref{q:eq:lift} is
$\det(\mu I_3-M_q)=(\mu+2\gamma/3)^3$.
For $N_M=M_q+2\gamma I_3/3$,
$\operatorname{rank}N_M=2$,
$\operatorname{rank}N_M^2=1$, and $N_M^3=0$.
Thus the coefficient dynamics has one size-three Jordan block at
$\mu_{\rm EP}=-2\gamma/3$.
The parameter set is related by a basis rotation to the known
third-order configuration of Ref.~\cite{lin2025}; here the
degeneracy is diagnosed in the NMQSD coefficient matrix itself.
Appendix~\ref{q:app:benchmark} gives the explicit basis rotation and
exact benchmark derivation.

\begin{figure*}[tbp]
 \centering
 \includegraphics[width=\textwidth]{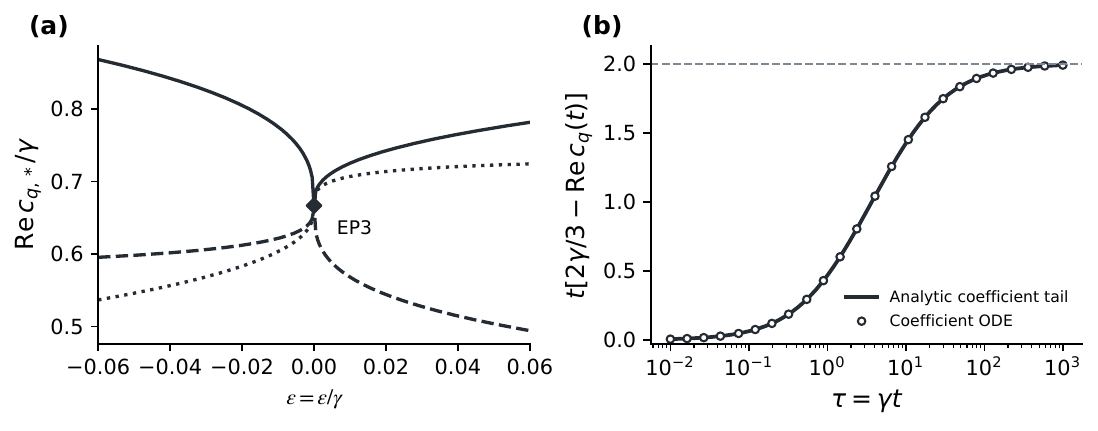}
 \caption{EP3 obtained from the NMQSD coefficient equations.
 (a) Real parts of the stationary contraction
 $c_{q,*}/\gamma=-\mu/\gamma$ under
 $H_{1,11}\mapsto H_{1,11}+\epsilon$.
 The branches are obtained from the coefficient lift
 in Eq.~\eqref{q:eq:lift}; the coalescence is fixed by the
 exact size-three Jordan block.
 (b) The ordinate $t[2\gamma/3-\operatorname{Re}c_q(t)]$
 approaches $2$. The analytic expression
 in Eq.~\eqref{q:eq:cEP} is compared with direct integration
 of the coefficient ordinary differential equation (ODE),
 Eq.~\eqref{q:eq:riccati}, using DOP853.
 This long-time comparison concerns coefficient dynamics.}
 \label{q:fig:structure}
\end{figure*}

With dimensionless time $\tau=\gamma t$, the prescribed
coefficient solution is
\begin{align}
 q(t)&=e^{-2\tau/3}
       \left(1+\frac{2\tau}{3}+\frac{2\tau^2}{27}\right),
 \label{q:eq:qEP}\\
 c_q(t)&=\gamma\left[
 \frac23-\frac{2/3+4\tau/27}{1+2\tau/3+2\tau^2/27}\right].
 \label{q:eq:cEP}
\end{align}
The chart remains finite because $q(t)>0$ for $t\ge0$.
Its asymptotic contraction is
$c_q=2\gamma/3-2/t+O((\gamma t^2)^{-1})$.
Figure~\ref{q:fig:structure} displays the stationary branches
and this coefficient tail.

For the local frequency shift
$H_{1,11}\mapsto H_{1,11}+\epsilon$, define
$\varepsilon=\epsilon/\gamma$ and
$\zeta=(\mu+2\gamma/3)/\gamma$.
The exact dimensionless coefficient polynomial is
\begin{equation}
 p_M(\zeta)=\zeta^3-i\varepsilon\zeta^2+
       \varepsilon(1/6+i/3)\zeta-\varepsilon/9 .
 \label{q:eq:cubic}
\end{equation}
It gives $\zeta^3\sim\varepsilon/9$ and hence a cube-root
splitting of the coefficient branches. The characteristic
polynomial and the Jordan-rank test specify the EP without
assuming its visibility in a particular state variable.

\subsection{Stochastic-state propagation and the exact NMME}
Let $\mathsf G(t)$ propagate the excitation amplitude of the
linear NMQSD state. Its system-excitation block is deterministic
in this sector:
\begin{equation}
 \dot{\mathsf G}
 =(-iH_1-\kappa^\dagger\bm F_q)\mathsf G,\qquad
 \mathsf G(0)=I_{N_q}.
 \label{q:eq:FG}
\end{equation}
The Gaussian history enters the zero-excitation amplitude.
Consequently, noise-independent coefficients do not imply
noise-independent stochastic kets.

The deterministic operator row factors out of the projector average in Eq.~\eqref{fw:nmme}: $\mathbb E_0[\bar O_t\Pi_t]=\bar O_t\rho_t$. This is the noise-independent case of the quantum-trajectory construction in Ref.~\cite{chen2014}. Thus, on the original retained system space,
\begin{equation}
 \dot\rho_t=-i[H_S,\rho_t]
       +[L,\rho_t\bar O_t^\dagger]
       +[\bar O_t\rho_t,L^\dagger].
 \label{q:eq:NMME}
\end{equation}
The time dependence of $\bar O_t$ retains the full exponential memory. Equation~\eqref{q:eq:NMME} is exact in the stated sector, without a Markov or weak-coupling approximation. Appendix~\ref{q:app:prior} gives its block reduction and integrated state map. Noise-dependent sectors in the full multiqubit problem require the additional averages treated in Ref.~\cite{chen2014}.

Write $\rho_{\mathrm{ee}}$ for the system excitation block,
$\rho_{\mathrm{e}G}=(\rho_{1G},\ldots,\rho_{N_qG})^T$ for
its coherence with $\ket{G_S}$, and $\rho_{GG}$ for the
zero-excitation population.
Inserting the explicit $L$ and $\bar O_t$ into
Eq.~\eqref{q:eq:NMME} gives
\begin{equation}
 \begin{aligned}
 \dot\rho_{\mathrm{e}G}
 &=(-iH_1-\kappa^\dagger\bm F_q)\rho_{\mathrm{e}G},\\
 \dot\rho_{\mathrm{ee}}
 &=(-iH_1-\kappa^\dagger\bm F_q)\rho_{\mathrm{ee}}\\
 &\quad+\rho_{\mathrm{ee}}(iH_1-\bm F_q^\dagger\kappa),\\
 \dot\rho_{GG}
 &=\bm F_q\rho_{\mathrm{ee}}\kappa^\dagger+
       \kappa\rho_{\mathrm{ee}}\bm F_q^\dagger .
 \end{aligned}
 \label{q:eq:rho_blocks}
\end{equation}
For a normalized initial excitation $u_0$, these equations imply
\begin{equation}
 \begin{aligned}
 \rho_t&=\ket{\mathsf G(t)u_0}\bra{\mathsf G(t)u_0}\\
 &\quad+[1-\|\mathsf G(t)u_0\|^2]\ket{G_S}\bra{G_S}.
 \end{aligned}
 \label{q:eq:density}
\end{equation}
For the coherent preparation
$(\ket{G_S}+\ket{u_0})/\sqrt2$, they instead give
\begin{equation}
 \rho_{\mathrm{e}G}(t)=\tfrac12\mathsf G(t)u_0 .
 \label{q:eq:coherence}
\end{equation}
This is an unconditional system coherence obtained from the
NMME. The reference superposition makes the readout linear in
$\mathsf G$, whereas populations involve amplitude products.
These distinct maps explain why one coefficient degeneracy need
not produce the same time dependence in every observable.

\subsection{Internal memory coordinates and pole transfer}
Define the row $\bm W=\bm F_q\mathsf G$. Its derivative,
using Eqs.~\eqref{q:eq:riccati} and \eqref{q:eq:FG}, is
\[
 \dot{\bm W}=A_0\kappa\mathsf G-d\bm W,\qquad \bm W(0)=0.
\]
Equivalently,
$\bm W(t)=A_0\int_0^t e^{-d(t-s)}\kappa\mathsf G(s)\,ds$.
It is a mathematical coordinate for the amplitude history.
The coupled propagation is therefore
\begin{equation}
 \begin{aligned}
 \frac{d}{dt}\binom{\mathsf G}{\bm W}
 &=\mathsf A_q\binom{\mathsf G}{\bm W},\qquad
 \binom{\mathsf G(0)}{\bm W(0)}=\binom{I_{N_q}}{0},\\
 \mathsf A_q&=
 \begin{pmatrix}
 -iH_1&-\kappa^\dagger\\
 A_0\kappa&-d
 \end{pmatrix}.
 \end{aligned}
 \label{q:eq:state_memory}
\end{equation}
This linear equation is derived from the NMQSD coefficients and
state amplitudes. It also supplies a global continuation when
a finite coefficient chart fails.

The coefficient and state-memory matrices satisfy
\begin{equation}
 \begin{aligned}
 M_q&=J(-\mathsf A_q^T-dI_{N_q+1})J,\\
 J&=\operatorname{diag}(I_{N_q},-1),\qquad
 \mu=-\lambda-d ,
 \end{aligned}
 \label{q:eq:similarity}
\end{equation}
where $\lambda$ is an eigenvalue of $\mathsf A_q$.
Their Jordan orders therefore coincide, although their eigenvalues and late-time ordering differ. Both matrices have been obtained from the same NMQSD coefficients and state map.
For $\operatorname{Re}\lambda<0$, the matrix
$\lambda I+iH_1$ is invertible, and the corresponding
eigenspace of $\mathsf A_q$ is one dimensional.
The prescribed coefficient initial condition retains the full pole
order because the Laplace transform of $q(t)$, with complex
frequency $s$, obeys
\begin{equation}
 \widetilde q(s)=e_a^T(sI-M_q)^{-1}e_a
 =\frac{\det[(s+d)I-iH_1^T]}{\det(sI-M_q)}
 \label{q:eq:qres}
\end{equation}
has a nonzero numerator at $s=-\lambda-d$.
A positive weighted-norm identity in
Appendix~\ref{q:app:proofs} shows that
$\operatorname{Re}\lambda\le0$ and that modes on the imaginary
axis are dark and semisimple. Thus a defective pole in this
class belongs to the decaying sector just considered.

If a reached order-$r$ pole $\mu_*$ is strictly dominant in real
part, the common projective criterion gives
\begin{equation}
 c_q(t)=-\mu_*-\frac{r-1}{t}+O(t^{-2}).
 \label{q:eq:tail}
\end{equation}
The affine map implies
\begin{equation}
 \operatorname{Re}\mu=-\operatorname{Re}\lambda-\gamma .
 \label{q:eq:reverse}
\end{equation}
The dominant coefficient pole therefore corresponds to faster
amplitude decay. Equal-real-part poles or a reached but
subleading EP require separate asymptotic analysis.

Let $U_q(t)=e^{\mathsf A_qt}$, whose system block is
$\mathsf G(t)$.
The inverse-transpose relation following from
Eq.~\eqref{q:eq:similarity} gives
\begin{equation}
 q(t)=e^{i\operatorname{tr}(H_1)t}\det\mathsf G(t).
 \label{q:eq:det}
\end{equation}
On an invertible chart, $\bm F_q=\bm W\mathsf G^{-1}$ and
$\bm f_q(t,s)=\kappa\mathsf G(s)\mathsf G(t)^{-1}$.
An isolated order-$m$ zero of $q$ at $t_0$ produces
$c_q\sim-m/(t-t_0)$, while the global state propagation remains
regular. Such coefficient-coordinate singularities are
compatible with regular reduced dynamics
\cite{q_Andersson2007}. They are distinct from spectral EPs;
Appendix~\ref{q:app:controls} separates chart poles, dark
degeneracies, and subleading EPs.

\begin{figure*}[tbp]
 \centering
 \includegraphics[width=\textwidth]{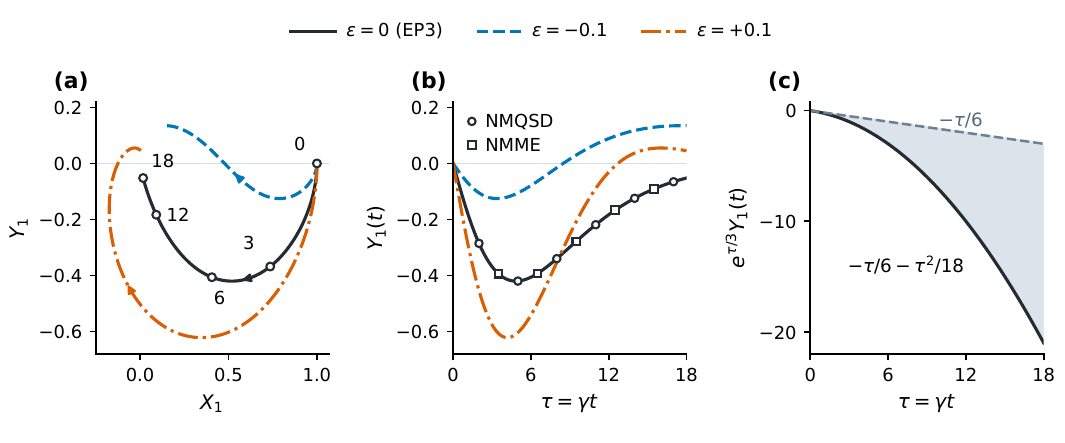}
 \caption{Unconditional coherence from NMQSD and the exact NMME.
 The initial system state is $(\ket{G_S}+\ket1)/\sqrt2$.
 Parameters are those of the coefficient EP3,
 Eq.~\eqref{q:eq:params}, with controls
 $H_{1,11}\mapsto H_{1,11}+\epsilon$ at
 $\varepsilon=\epsilon/\gamma=\pm0.1$.
 Here $X_1=2\operatorname{Re}\rho_{1G}$ and
 $Y_1=2\operatorname{Im}\rho_{1G}$ are the phase-referenced
 system-coherence quadratures.
 (a) Their $(X_1,Y_1)$ trajectories.
 Arrows indicate increasing time; circles on the EP curve mark
 $\tau=\gamma t=0,3,6,12,18$.
 (b) Raw $Y_1(t)$. The state-response curves are compared on the
 EP trace with NMQSD coefficient propagation (circles) and
 direct NMME density propagation (squares).
 (c) Compensation gives
 $e^{\tau/3}Y_1(t)=-\tau/6-\tau^2/18$.
 The dashed linear term and shaded quadratic contribution expose
 the nonzero third-order coherence pole; the raw signal decays.
 Here $A_0/\gamma^2=1$, $\Omega/\gamma=0$,
 and the laboratory carrier is $\omega_0/\gamma=2$.
 Numerical settings and discrepancies are given in
 Appendix~\ref{q:app:numerics}.}
 \label{q:fig:coherence}
\end{figure*}

The system readout must still be checked. For
\begin{equation}
 \begin{aligned}
 D_q(z)&=zI+iH_1,\\
 h_q(z)&=z+d+A_0\kappa D_q(z)^{-1}\kappa^\dagger ,
 \end{aligned}
 \label{q:eq:schur_functions}
\end{equation}
the Laplace transform of $\mathsf G(t)$ is the system block of the
state-memory resolvent at complex frequency $z$:
\begin{equation}
 \widetilde{\mathsf G}(z)
 =D_q^{-1}-A_0D_q^{-1}\kappa^\dagger h_q^{-1}\kappa D_q^{-1}.
 \label{q:eq:schur}
\end{equation}
At a decaying order-$r$ EP, $h_q$ has an order-$r$ zero at
$\lambda$. The highest Laurent coefficient in this system block is
\begin{equation}
 (R_r)_{\mathrm{ee}}=
 -\frac{A_0r!}{h_q^{(r)}(\lambda)}
 D_q(\lambda)^{-1}\kappa^\dagger\kappa D_q(\lambda)^{-1}
 \ne0 .
 \label{q:eq:residue}
\end{equation}
Both factors are nonzero for the nonzero coupling required by
the EP. Consequently, every defective decaying mode contributes $t^{r-1}e^{\lambda t}$ to at least one system coherence for a suitable preparation. This establishes transfer of the full pole order to an unconditional readout. It does not require every quadrature to retain the term, as the EP3 quadratures in Fig.~\ref{q:fig:coherence} show.

\subsection{EP3 in system quadratures}
For the coefficient EP3 in Eq.~\eqref{q:eq:params},
the state-memory eigenvalue is $\lambda_{\rm EP}=-\gamma/3$.
Prepare $(\ket{G_S}+\ket1)/\sqrt2$.

The Hermitian observables
\begin{equation*}
 \begin{aligned}
  \widehat X_1&=\ket{G_S}\bra1+\ket1\bra{G_S},\\
  \widehat Y_1&=-i\ket{G_S}\bra1+i\ket1\bra{G_S}
 \end{aligned}
\end{equation*}
have means
$X_1=2\operatorname{Re}\rho_{1G}$ and
$Y_1=2\operatorname{Im}\rho_{1G}$.
The NMQSD propagation and Eq.~\eqref{q:eq:coherence} give
(see Appendix~\ref{q:app:benchmark} for the amplitude derivation)
\begin{align}
 X_1(t)&=e^{-\tau/3}(1+\tau/3),\nonumber\\
 Y_1(t)&=-e^{-\tau/3}(\tau/6+\tau^2/18).
 \label{q:eq:quadratures}
\end{align}
Equivalently, the Laplace transform of $C_1=\rho_{1G}$ is
\begin{equation}
 \begin{aligned}
 \widetilde C_1(s)
 &=\frac{1}{2(s+\gamma/3)}
   +\frac{(2-i)\gamma}{12(s+\gamma/3)^2}\\
 &\quad-\frac{i\gamma^2}{18(s+\gamma/3)^3}.
 \end{aligned}
 \label{q:eq:Cpole}
\end{equation}
The same preparation thus gives a quadratic Jordan term in $Y_1$ and only a linear term in $X_1$. The difference is a readout projection, not a change in the underlying EP.
A common laboratory carrier $\omega_0=2\gamma$ gives
$H_{1,\rm lab}=H_1+\omega_0I_2$ and
$\Omega_{\rm lab}=\omega_0$.
The coefficient matrix is unchanged, while
$C_{1,\rm lab}=e^{-i\omega_0t}C_1$; a phase reference selects
the quadratures plotted in Fig.~\ref{q:fig:coherence}.

The same preparation also allows the raw local selection
\begin{equation}
 X_2(t)=2\operatorname{Re}\rho_{2G}
       =-\frac{\sqrt3}{27}\tau^2e^{-\tau/3}.
 \label{q:eq:X2pure}
\end{equation}
On the retained sector this is the mean of $I\otimes\sigma_x$.
It isolates the quadratic term without compensation and reaches
a maximum magnitude of approximately $0.31$ at $\tau=6$.
Appendix~\ref{q:app:numerics} discusses phase selection and
finite-signal diagnostics. The exact polynomial and the nonzero residue establish this signature; a short-time quadratic fit alone would not certify an EP.

\subsection{Recovering coefficients from system readouts}
For separate preparations $(\ket{G_S}+\ket j)/\sqrt2$,
the phase-calibrated unconditional coherences determine each column:
\begin{equation}
 \mathsf G_{ij}(t)=2\rho_{iG}^{(j)}(t)
                 =X_i^{(j)}(t)+iY_i^{(j)}(t).
 \label{q:eq:coherenceG}
\end{equation}
On an invertible chart, Eq.~\eqref{q:eq:FG} gives
\begin{align}
 \bm F_q&=-\frac{\kappa(\dot{\mathsf G}\mathsf G^{-1}+iH_1)}
                     {\kappa\kappa^\dagger},
 \label{q:eq:inverseF}\\
 c_q&=-\operatorname{tr}(\dot{\mathsf G}\mathsf G^{-1})
       -i\operatorname{tr}H_1 .
 \label{q:eq:inversec}
\end{align}
The row reconstruction requires known $H_1$ and known nonzero
$\kappa$ in a fixed normalization and phase convention; the
trace formula needs only known $\operatorname{tr}H_1$.
Under a common laboratory rotation,
$\mathsf G_{\rm lab}=e^{-i\omega_0t}\mathsf G_{\rm rot}$,
$H_{1,\rm lab}=H_{1,\rm rot}+\omega_0I_{N_q}$, and
$\Omega_{\rm lab}=\Omega+\omega_0$. The determinant-phase derivative
and the trace shift cancel in Eq.~\eqref{q:eq:inversec}.
This application of evolution-map inversion
\cite{q_Andersson2007} identifies the integrated coefficients
of the specified NMQSD representation.
Different readouts require separately prepared ensembles.
Differentiation and small singular values of $\mathsf G$ can
amplify errors even while the physical state stays regular.
For the two-qubit example, both basis preparations are needed
to recover the complete map; the selected $X_2$ record alone
does not determine the full coefficient row.
The interval reconstruction and conditioning checks are given in Appendix~\ref{q:app:inverse}.

\section{Position-coupled oscillator in an optomechanical sector}\label{sec:optomech}

\subsection{Fixed-sector closure and Gaussian propagation}
\label{o:sec:model}

Position coupling requires a noise-dependent $O$ operator even for a harmonic oscillator. This makes the model complementary to the qubits: a coefficient branch that is subleading in the causal solution can still enter the state through its width, and a narrow conditional state can become broad after averaging over its centers. We retain the full noise-history term to follow both effects.

We use dimensionless mechanical quadratures with $[Q,P]=\ii$ and vacuum covariance $\Id_2/2$. The system Hamiltonian and bath coupling operator are
\begin{equation}
H_S=\frac{\nu}{2}(Q^2+P^2)+\eta NQ+H_N,\qquad L=\ell Q.
\label{o:eq:H}
\end{equation}
For the optomechanical realization, let $a_c$ be a cavity annihilation operator, $N=a_c^\dagger a_c$, and $H_N=\omega_cN$. The mechanical position is $Q$, and $\eta NQ$ is the radiation-pressure interaction in the chosen sign convention \cite{aspelmeyer2014}. The photon-number dependence of the mechanical force is the familiar nonlinear optomechanical coupling used in state-preparation models \cite{bose1997}. Here $[N,Q]=[N,P]=0$ and the photon number is conserved. We restrict the system preparation to one photon-number sector, $N\ket n=n\ket n$. No cavity drive or optical loss is included in this fixed-sector model.

We fix $n>0$ and write its constant energy as $E_n$. The oscillator frequency $\nu$, coupling scale $\ell$, and force coefficient $\eta$ are positive. Within this sector, radiation pressure is an affine force. The coefficient core and its exceptional-point locus below are independent of $\eta$ and $n$; the result does not characterize driven cavity--mechanical hybrid-mode exceptional points.

The initial oscillator state is a normalized pure Gaussian, independent of the vacuum Gaussian bath. We use the common correlation in Eq.~\eqref{fw:kernel}, now with $A_0,\gamma,\Omega>0$, and define $B=A_0\ell^2$. Position coupling retains both rotating and counter-rotating terms, placing the oscillator in the quantum Brownian-motion setting \cite{hu1992,strunz2004}. The exact NMQSD and consistency equations are Eqs.~\eqref{fw:qsd} and \eqref{o:eq:consistency}.

For Eq.~\eqref{o:eq:H}, the operator algebra closes on $Q,P,N,\Id$. The noise-complete oscillator $O$ operator of Ref.~\cite{dgs1998} extends to this conserved affine sector. In particular, $[\ell Qz_t^*,g_OP]=\ii\ell g_Oz_t^*\Id$ requires the same scalar noise term:
\begin{equation}
\begin{aligned}
O(t,s,z^*)&=f(t,s)Q+g_O(t,s)P+r_O(t,s)N\\
&\quad+\ii\int_0^t j(t,s,u)z_u^*\dd u\,\Id.
\end{aligned}
\label{o:eq:O}
\end{equation}
Writing $\bar O_t=FQ+GP+RN+\ii\Xi_t\Id$, define $F(t)=\int_0^t\alpha(t,s)f(t,s)\dd s$, with $G(t)$ and $R(t)$ given by the same integral of $g_O(t,s)$ and $r_O(t,s)$. The integrated scalar noise history is $\Xi_t=\int_0^t\mathcal J(t,u)z_u^*\dd u$, where $\mathcal J(t,u)=\int_0^t\alpha(t,s)j(t,s,u)\dd s$. Its deterministic bath integral $J(t)=\int_0^t\alpha(t,u)\mathcal J(t,u)\dd u$ enters $\dot F$ through the functional-derivative contribution $-\ii\ell J$. The $(F,G,J)$ equations are the exponential-kernel oscillator equations of Ref.~\cite{dgs1998}, here with the complex decay $d$ and the affine coefficient $R$:
\begin{align}
\dot F&=A_0\ell-dF+\nu G-\ii\ell FG-\ii\ell J,\nonumber\\
\dot G&=-dG-\nu F-\ii\ell G^2,\nonumber\\
\dot J&=A_0\ell G-2dJ-\ii\ell GJ,\nonumber\\
\dot R&=-dR+\eta G-\ii\ell GR,
\label{o:eq:coeff}
\end{align}
with $F(0)=G(0)=J(0)=R(0)=0$. Appendix~\ref{o:app:O} gives the two-time equations and both noise boundaries. This closure uses no weak-coupling, rotating-wave, or noise-order truncation. A Gaussian system preparation is needed for the state representation below, but not for the operator closure. The scalar $\Xi_t$ still multiplies $Q$ in Eq.~\eqref{fw:qsd}, so it changes the displacement rather than merely the overall phase.

The deterministic core $(F,G,J)$ fixes the coefficient spectrum, while $R$ carries the affine force. The scalar history remains necessary for the displacement and for the operator average.

Write the unnormalized stochastic wavefunction as $\psi_t(x;z^*)=\exp[-a_tx^2/2+b_tx+c_{\psi,t}]$, with $a_R=\Rea a>0$ and $a_I=\Ima a$. Substitution into Eq.~\eqref{fw:qsd} yields
\begin{align}
\dot a&=\ii\nu(1-a^2)+2\ell F+2\ii\ell Ga,\nonumber\\
\dot b&=(\ii\ell G-\ii\nu a)b+\ell z_t^*-\ii\ell\Xi-n(\ii\eta+\ell R),\nonumber\\
\dot c_\psi&=\frac{\ii\nu}{2}(b^2-a)-\ii E_n,\nonumber\\
\dot\Xi&=\ell Gz_t^*-(d+\ii\ell G)\Xi,\qquad \Xi(0)=0.
\label{o:eq:Gaussian}
\end{align}
We order the quadratures as $r=(Q,P)^T$ and use the mean $m=\langle r\rangle$ and symmetrized covariance $V_{ij}=\langle\{r_i-m_i,r_j-m_j\}\rangle/2$ for a normalized state \cite{weedbrook2012}. For the normalized stochastic Gaussian, these are
\begin{align}
m_\psi&=T_a\binom{\Rea b}{\Ima b},\qquad
T_a=\begin{pmatrix}1/a_R&0\\-a_I/a_R&1\end{pmatrix},\label{o:eq:centroid}\displaybreak[0]\\
V_\psi&=\frac{1}{2a_R}\begin{pmatrix}1&-a_I\\-a_I&|a|^2\end{pmatrix}.
\label{o:eq:pureV}
\end{align}
The width is deterministic, while the displacement responds linearly to the noise history. The scalar wavefunction coefficient $c_\psi$ determines the squared norm used in the average, even though it cancels from the normalized moments in Eqs.~\eqref{o:eq:centroid} and \eqref{o:eq:pureV}. It is distinct from the scalar-noise sector $\Xi$ of the $O$ operator. The sector energy $E_n$ contributes only a phase.

\subsection{Coefficient EPs and the two-root state response}
\label{o:sec:branches}
\subsubsection{The causal coefficient solution}
The nonlinear subsystem $(F,G,J)$ has a four-dimensional linear representation:
\begin{equation}
\dot v={M_o} v,\quad v(0)=e_4,\quad
{M_o}=\begin{pmatrix}-d&\nu&-\ii\ell&A_0\ell\\-\nu&-d&0&0\\0&A_0\ell&-2d&0\\0&\ii\ell&0&0\end{pmatrix}.
\label{o:eq:M}
\end{equation}
Here $(F,G,J)=(x_F,x_G,x_J)/q$, $v=(x_F,x_G,x_J,q)^T$, and $e_4=(0,0,0,1)^T$ is the fourth unit vector.
The characteristic polynomial and its shifted discriminant are
\begin{align}
P_a(\mu)&=\mu(\mu+2d)[(\mu+d)^2+\nu^2]+2\ii B\nu d,\label{o:eq:Pa}\\
\Delta_a&=(\nu^2+d^2)^2-8\ii B\nu d.\label{o:eq:Delta}
\end{align}
Their positive-parameter EP branch is
\begin{align}
\Omega_a^2&=\frac{\nu^2-\gamma^2+2\sqrt{\nu^4+\nu^2\gamma^2+\gamma^4}}{3},\nonumber\\
B_a&=\frac{\Omega_a(\nu^2+\gamma^2-\Omega_a^2)}{2\nu}.
\label{o:eq:locus}
\end{align}
In this oscillator section, $\lambda=B/B_a$ denotes the dimensionless coupling ratio. Away from coalescence, define the two shifted roots with positive real parts,
\begin{equation}
w_{1,2}=\sqrt{\frac{d^2-\nu^2\pm\sqrt{\Delta_a}}{2}},
\qquad \Rea w_{1,2}>0.
\label{o:eq:shiftedroots}
\end{equation}
The four eigenvalues $\mu=-d\pm w_1,-d\pm w_2$ give four finite algebraic fixed points away from coalescence. Figure~\ref{o:fig:branches} follows the two positive-shift branches, $G_*^{(j)}=(-d+w_j)/(\ii\ell)$. The causal trajectory $G(t)=\dot q(t)/(\ii\ell q(t))$ approaches the branch with larger $\Rea w_j$ for simple roots of unequal real parts, provided the prescribed initial vector reaches that component and the ratio chart is regular. Other reached components remain at finite age. State reconstruction also requires the Gaussian width.

At the EP, put $w^2=(d^2-\nu^2)/2$ with $\Rea w>0$. Each of $\mu_\pm=-d\pm w$ has algebraic multiplicity two and geometric multiplicity one. The initial vector $e_4$ reaches the leading Jordan chain, giving
\begin{equation}
G(t)=\frac{\mu_+}{\ii\ell}+\frac{1}{\ii\ell t}
+O(t^{-2})+O\!\left[\mathrm{poly}(t)e^{-2\Rea w\,t}\right].
\label{o:eq:Gtail}
\end{equation}
Appendix~\ref{o:app:spectral} proves the positive branch and its causal reachability. A zero of $q$ is a pole of this coefficient chart; the homogeneous evolution in Eq.~\eqref{o:eq:M} remains regular.

\subsubsection{Width selection and displacement response}

In the EP neighborhood, label the simple roots locally so that $\Rea w_1>\Rea w_2>0$, and set $\mu_1=w_1-d$. The selected coefficient solution approaches
\begin{equation}
G_\infty=\frac{\mu_1}{\ii\ell},\qquad
F_\infty=\frac{\ii\mu_1(\mu_1+d)}{\nu\ell}.
\label{o:eq:branchlimit}
\end{equation}
Since $w_1^2+w_2^2=d^2-\nu^2$, the stable normalizable width continued from the EP is
\begin{equation}
a_\infty=\frac{\ii(d-w_1-w_2)}{\nu}.
\label{o:eq:branchwidth}
\end{equation}
Its linearization rate is $-2w_2$; the other width root has rate $+2w_2$. Thus the companion coefficient root appears through the width, while the selected root remains in the noise-history channel. Comparing histories at the same width preparation and the same nonzero affine force removes the common force term. For a history input at time $s$ and endpoint $T$, let $\theta=T-s\geq0$ be the lag. In the long-preparation-age limit $T\to\infty$ at fixed $\theta$, the variations satisfy
\begin{equation}
\begin{aligned}
\frac{\dd}{\dd\theta}\binom{\delta b}{\delta\Xi}
&=A_\infty\binom{\delta b}{\delta\Xi}+B_\infty\delta z^*,\\
A_\infty&=\begin{pmatrix}-w_2&-\ii\ell\\0&-w_1\end{pmatrix},\qquad
B_\infty=\binom{\ell}{-\ii\mu_1}.
\end{aligned}
\label{o:eq:branchresponse}
\end{equation}

For zero initial response, the Laplace transform in $\theta$, with complex variable $p$, gives the transfer function
\begin{equation}
H_{bz}(p)=\frac{\ell(p+d)}{(p+w_1)(p+w_2)}.
\label{o:eq:transfer}
\end{equation}
The direct noise input and the history stored in $\Xi$ both reach $b$. For $B\nu>0$ and $d\ne0$, neither root equals $d$, so both residues are nonzero away from coalescence.

Combining the two poles before taking their coalescence gives a regular expression for the complete kernel,
\begin{equation}
\begin{aligned}
h(\theta)&=\ell e^{-\bar w\theta}\bigl[\cosh(\epsilon_w\theta)
 +(d-\bar w)\theta\sinhc(\epsilon_w\theta)\bigr],\\
\bar w&=(w_1+w_2)/2,\qquad \epsilon_w=(w_1-w_2)/2,
\end{aligned}
\label{o:eq:nearEPkernel}
\end{equation}
where $\sinhc x=\sinh(x)/x$ and $\sinhc0=1$. At the EP,
\begin{equation}
h_{\rm EP}(\theta)=\ell[1+(d-w)\theta]e^{-w\theta}.
\label{o:eq:EPkernel}
\end{equation}
The nonzero coefficient $\ell(d-w)$ gives a double pole in the stochastic displacement response. The polynomial--exponential form is the familiar time-domain consequence of a higher-order pole \cite{fuchs2014}; here its residue is fixed by the causal preparation and the specified history-to-displacement map. Through $T_a$ in Eq.~\eqref{o:eq:centroid}, this kernel quantifies how a variation $\delta z^*$ of the NMQSD Bargmann history changes the normalized conditional centroid. A force susceptibility requires an applied-force input, while bath-averaged two-time correlations require their own averaging construction. Such correlations can be obtained by propagating paired trajectories \cite{carballeira2021}; the optomechanical treatment in Ref.~\cite{shi2026} uses a common phononic history with optical no-jump evolution.

\begin{figure*}[tbp]
\centering
\includegraphics[width=\textwidth]{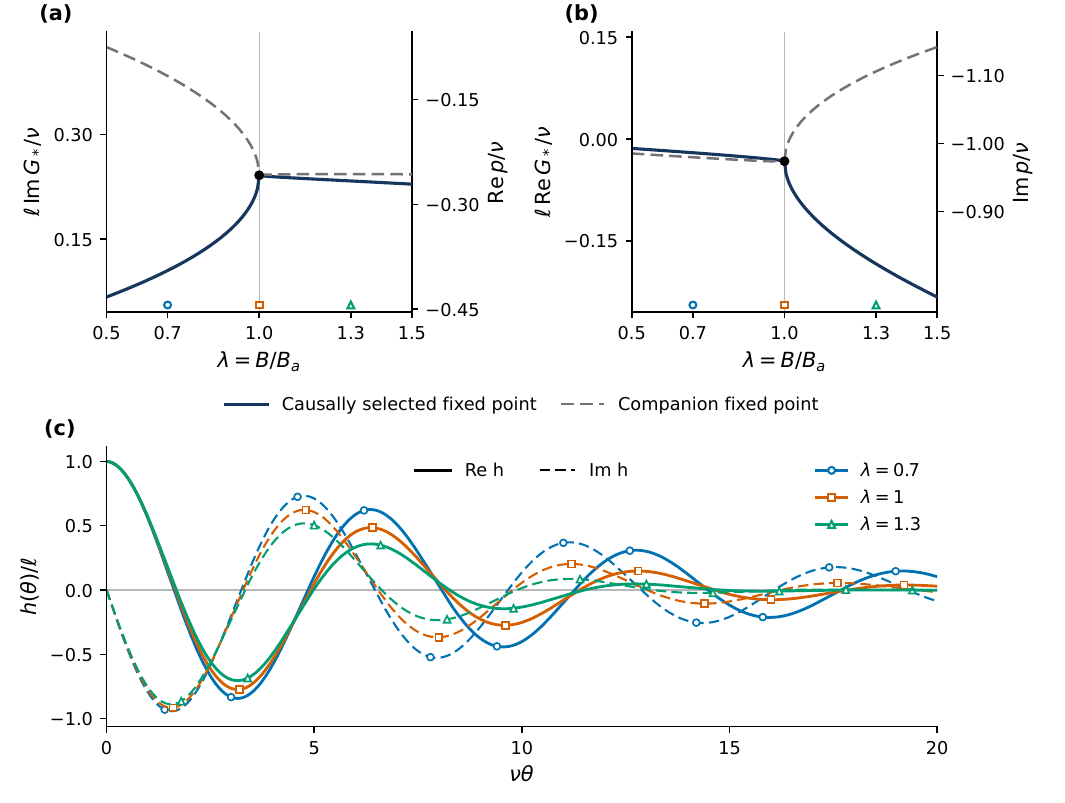}
\caption{Coefficient branches and the two channels of the stochastic-state response. (a),(b) The two positive-shift fixed-point branches $G_*^{(j)}=(-d+w_j)/(\ii\ell)$ along $\lambda=B/B_a$ at $\Omega=\Omega_a$ coalesce at $\lambda=1$. The subscript $*$ denotes a fixed point. The solid branch is selected by the causal zero initial condition; the dashed companion still enters the state response through the second root. Right axes show the corresponding response poles $p=-w_{1,2}$ on the same curves. (c) Real (solid) and imaginary (dashed) parts of the full kernel $h(\theta)$ at $\lambda=0.7,1,1.3$, identified respectively by sparse circles, squares, and triangles. At coalescence, the response contains both $e^{-w\theta}$ and $\theta e^{-w\theta}$. The horizontal axis is the dimensionless lag $\nu\theta$. This is the fixed-lag response after a long preparation age, with $0\leq\nu\theta\leq20$, rather than the finite-age evolution shown in Sec.~\ref{o:sec:results} and Appendix~\ref{o:app:validationfigures}. Parameters are $\nu=\ell=1$, $\gamma=0.5$, $\eta=0.25$, and $n=1$.}
\label{o:fig:branches}
\end{figure*}

Along $\Omega=\Omega_a$, the splitting in the coupling ratio is
\begin{equation}
w_1-w_2=\frac{\sqrt{-8\ii B_a\nu d(\lambda-1)}}{2w}
+O(|\lambda-1|^{3/2}).
\label{o:eq:splitting}
\end{equation}
The roots split as a square root, whereas Eq.~\eqref{o:eq:nearEPkernel} is analytic in $\lambda$ near $\lambda=1$ at every fixed finite lag. Figure~\ref{o:fig:branches} shows both the branch geometry and the damped complex response.

The width has a different late-time consequence. On the regular normalizable branch at the EP, for $0<\gamma<\nu$,
\begin{equation}
a_* =\frac{\ii(d-2w)}{\nu},\qquad
a(t)-a_*=O\!\left(t^2e^{-2\Rea w\,t}\right).
\label{o:eq:width}
\end{equation}
The purely algebraic coefficient tail cancels from the width equation; polynomially modified exponential corrections remain. The interval $0<\gamma<\nu$ places the bath correlation decay rate below the oscillator frequency and is the working interval used here together with the regular normalizable branch condition. Appendix~\ref{o:app:width} proves that the stationary width has $\Rea a_*>0$ for all positive $\gamma,\Omega,\nu$; this algebraic positivity alone does not establish finite-time normalizability. On the regular branch, preservation of the mean squared norm in linear NMQSD excludes the nonnormalizable asymptotic solution and gives Eq.~\eqref{o:eq:width}. The finite-time reduced state requires the weighted average developed next.

\subsection{Complete noise average and finite-time state dynamics}
\label{o:sec:average}
The common average in Eq.~\eqref{fw:average} can be evaluated directly for the Gaussian state. The reference measure $\mathbb P_0$ averages unnormalized projectors, whereas the norm-weighted measure $\mathbb P_t$ averages normalized-state observables. Because $c_\psi$ determines that weight, it must be kept even when the desired observable depends only on $a$ and $b$.

\subsubsection{The Gaussian ensemble and its norm weight}
\begin{figure*}[tbp]
\centering
\includegraphics[width=\textwidth]{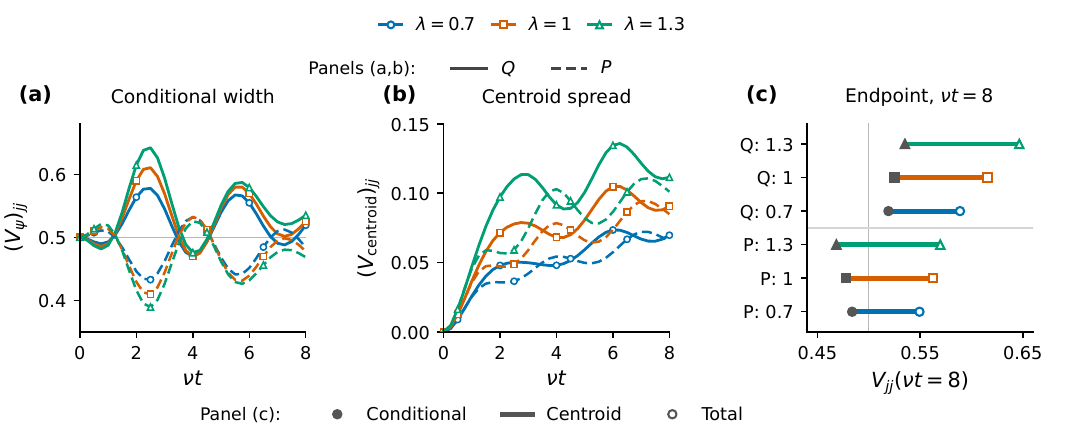}
\caption{Conditional width and centroid spread resolve the reduced-state variance. (a) conditional covariance $V_\psi$; (b) centroid covariance $V_{\rm centroid}=\operatorname{Cov}_{\mathbb P_t}(m_\psi)$; (c) their relation to the total variance at $\nu t=8$. In (a),(b), $j=Q,P$: solid and dashed lines denote $Q$ and $P$, respectively; all time-series markers are hollow. Color and marker shape identify the three fixed couplings. Each row in (c) identifies one quadrature and coupling; a gray filled symbol marks the conditional variance and a hollow symbol marks the total. Their horizontal separation equals the centroid contribution. The rows are discrete cases, with no interpolation between them. Faint reference lines mark the initial vacuum variance $1/2$. Across the three momentum rows, increasing $\lambda$ moves the conditional endpoint left and the total endpoint right: centroid spread outweighs conditional narrowing. Parameters and the common Gaussian initial state are specified in Sec.~\ref{o:sec:results}.}
\label{o:fig:decomposition}
\end{figure*}

Applying the same mean and symmetrized-covariance convention to the reduced state, the norm-weighted ensemble of the conditional moments in Eqs.~\eqref{o:eq:centroid} and~\eqref{o:eq:pureV} gives
\begin{equation}
\begin{aligned}
m&=\mathbb E_{\mathbb P_t}[m_\psi],\\
V&=V_\psi+V_{\rm centroid},\\
V_{\rm centroid}&=\Cov_{\mathbb P_t}(m_\psi).
\end{aligned}
\label{o:eq:totalvariance}
\end{equation}
Because the Gaussian width is independent of the noise labels, the first covariance contribution is the same for every normalized stochastic state. The second measures the spread of their centers under the norm-weighted measure. A reduction of a conditional quadrature variance therefore does not determine the corresponding reduced-state variance, as Fig.~\ref{o:fig:decomposition} illustrates for the preparations in Sec.~\ref{o:sec:results}.

For the numerical average, the finite Gaussian source in Appendix~\ref{o:app:numerics} makes the displacement affine and the scalar $c_\psi$ quadratic in the source coordinates. Appendix~\ref{o:app:Gaussian} evaluates the resulting Gaussian norm integral and moments, retaining $c_\psi$ and checking the absolute total weight $Z$ against its exact value one. Source refinement and agreement with the effective master equation are reported in Appendix~\ref{o:app:numerics}.

\subsubsection{Finite-time reduced-state dynamics}\label{o:sec:results}
We compare the coefficient EP and its two sides at fixed $\nu=\ell=1$, $\gamma=0.5$, $\eta=0.25$, $n=1$, and $\Omega=\Omega_a$. The three couplings $B=\lambda B_a$, with $\lambda=0.7,1,1.3$, share the normalized initial wavefunction $a(0)=1$, $b(0)=0$, $c_\psi(0)=-\log\pi/4$, so that $m(0)=0$ and $V(0)=\Id_2/2$. We follow the full nonautonomous equations over \mbox{$0\leq\nu t\leq8$}. The spectral coalescence is identified analytically by Eq.~\eqref{o:eq:locus}; these three trajectories describe the state evolution at the specified couplings.

The variance decomposition in Fig.~\ref{o:fig:decomposition} separates the width of a normalized stochastic pure state from the spread of its center in the norm-weighted ensemble. Panels (a) and (b) follow these two contributions on the same time grid as the total variances reported in Appendix~\ref{o:app:validationfigures}. Their sum is the reduced-state variance in Eq.~\eqref{o:eq:totalvariance}; a narrower conditional state therefore need not imply a narrower reduced state.

At the common endpoint $\nu t=8$, increasing the coupling across the three examples narrows the conditional momentum distribution, while the spread of its centers grows enough to raise the total momentum variance. At $\lambda=1$, for example, the conditional contribution is about $0.48$, below the initial vacuum value $1/2$, but a centroid contribution of about $0.08$ raises the total to $0.56$.

At finite times, the combined state can remain smooth in the parameters on a regular coefficient and averaging chart, even though the labeled roots split as a square root. The three fixed-coupling trajectories describe this finite-time averaged evolution; the analytic EP result concerns the coalescence of the two stochastic history-response channels and their nonzero Jordan term at long preparation age and fixed lag.

\subsection{The effective master equation}
Averaging Eq.~\eqref{fw:qsd} with $\mathcal A_t=\mathbb E_0[\bar O_t\Pi_t]$ gives the formal master equation \cite{strunz2004,chen2014}
\begin{equation}
\dot\rho_t=-\ii[H_S,\rho_t]+[L,\mathcal A_t^\dagger]-[L^\dagger,\mathcal A_t].
\label{o:eq:formalME}
\end{equation}
Following the paired operator-average construction of Strunz and Yu \cite{strunz2004}, resolve $\mathcal A_t$ into left and right actions, $\mathcal A_t=\sum_k(U_kX_k\rho_t+V_k\rho_tX_k)$, with $X_k\in\{Q,P,N\}$. On the regular averaging chart specified in Appendix~\ref{o:app:average}, the effective coefficients are
\begin{equation}
\mathfrak f=U_Q-V_Q^*,\quad \mathfrak g=U_P-V_P^*,\quad \mathfrak r=U_N-V_N^*.
\label{o:eq:effective}
\end{equation}
The coefficients $U_k,V_k$ are the bath-weighted integrals of the left/right components in Eq.~\eqref{o:eq:UV}. They retain scalar-noise correlations, so the effective coefficients generally differ from $F,G,R$. Appendix~\ref{o:app:average} derives the paired integral inverse for the present exponential kernel and affine sector, including its pointwise-conjugation convention and invertibility condition.

Thus $\Ima\mathfrak f$ changes the restoring force, $\Ima\mathfrak g$ the damping, and $\Ima\mathfrak r$ the nonzero affine force. The real parts of $\mathfrak f$ and $\mathfrak g$ determine momentum and cross diffusion.

In the sector $N=n$, the closed master equation is
\begin{align}
\dot\rho={}&-\ii[H_{\rm eff},\rho]-\ell\Rea\mathfrak f[Q,[Q,\rho]]\nonumber\\
&-\ell\Rea\mathfrak g[Q,[P,\rho]]
-\ii\ell\Ima\mathfrak g[Q,\{P,\rho\}],\label{o:eq:closedME}\\
H_{\rm eff}={}&\frac\nu2(Q^2+P^2)+\ell\Ima\mathfrak f\,Q^2\nonumber\\
&+(\eta n+2\ell n\Ima\mathfrak r)Q+E_n.
\label{o:eq:Heff}
\end{align}
The reduced-state moments defined above obey
\begin{equation}
\dot m=\mathsf A m+\mathbf b,\qquad
\dot V=\mathsf A V+V\mathsf A^T+\mathsf D,
\label{o:eq:moments}
\end{equation}
where
\begin{align}
\mathsf A&=\begin{pmatrix}0&\nu\\-\nu-2\ell\Ima\mathfrak f&-2\ell\Ima\mathfrak g\end{pmatrix},\nonumber\\
\mathbf b&=\binom{0}{-\eta n-2\ell n\Ima\mathfrak r},\nonumber\\
\mathsf D&=\begin{pmatrix}0&-\ell\Rea\mathfrak g\\-\ell\Rea\mathfrak g&2\ell\Rea\mathfrak f\end{pmatrix}.
\label{o:eq:ABD}
\end{align}
Physicality concerns the propagated state; the instantaneous diffusion matrix need not be positive semidefinite in this non-Markovian equation. For a Gaussian preparation, $m$ and $V$ specify the reduced Gaussian state. The accompanying effective-coefficient and full-state comparisons are given in Appendix~\ref{o:app:validationfigures}.

\section{Physical meaning and scope of the signatures}
\label{sec:synthesis}

The two models answer the same question at different stages of the reconstruction. In the qubits, the deterministic lowering row passes directly through the noise average, and a reference coherence is linear in the system amplitude. A nonzero highest-order resolvent residue therefore gives an unconditional Jordan contribution. In the oscillator, the scalar noise history survives in the state equation and in the operator average. The width combines both coefficient roots before the displacement is formed, and norm weighting adds the spread of centers to the conditional covariance.

Three selections remain distinct: the causal initial vector must reach a Jordan chain; dominance in the denominator projection determines whether that chain sets a coefficient's algebraic tail; the chosen state or observable must retain its residue. The qubit eigenvalue relation $\mu=-\lambda-d$ reverses coefficient dominance and amplitude-decay ordering, while the oscillator's causal coefficient approaches one branch whose companion remains in the normalizable width and displacement response.

The observable determines what a measured time trace can establish. In the qubit example, a calibrated phase reference selects an unconditional coherence with a quadratic--exponential term. Both basis preparations, known model parameters, and an invertible amplitude map then allow the full coefficient row to be reconstructed. The oscillator's double pole describes the conditional displacement response of the chosen NMQSD representation to a history variation at long preparation age and fixed lag. Applied-force susceptibilities and bath-averaged two-time correlations involve separate propagation problems \cite{carballeira2021,shi2026}. The finite-time variance comparison establishes the effect of complete averaging for the displayed parameters; it does not isolate an EP-specific enhancement.

Finite correlation time specifies a reservoir property, whereas operational non-Markovianity is assessed from the reduced dynamical map through information-backflow or divisibility criteria \cite{blp2009,rhp2010,rivas2014,chruscinski2022}. Their relation to a coefficient EP is model dependent, as is the occurrence of revivals in reaction-coordinate dynamics \cite{holst2026}. Likewise, an instantaneous time-local NMME generator has its own spectral structure, which must be determined separately from that of the coefficient lift.

The exact results apply to the stated closures and preparations. Multiple excitations introduce additional operator sectors \cite{chen2014}, while finite temperature extends the stochastic trajectories and their averages \cite{yu2004}. Stochastic hierarchies accommodate more general reservoir correlations \cite{suess2014}; their spectral and averaging analysis must be established for the enlarged hierarchy, with master-equation consistency and positivity checked for approximate closures \cite{shi2024}. In the oscillator calculation, conserved photon number and a pure Gaussian mechanical preparation specify the state family. Its long-age results are established on a regular normalizable branch in the chosen interval $0<\gamma<\nu$, and the explicit master equation uses a regular averaging chart. These conditions connect the coefficient EP to the particular state signature under study.

\section{Conclusions}
\label{sec:conclusions}

We have connected exceptional points of exact NMQSD coefficient equations to state dynamics by carrying the same operator closure through reconstruction and noise averaging. The coefficient lift identifies reached Jordan chains and their asymptotic contributions, while the state map and the readout determine which terms survive. This separation also distinguishes a spectral EP from a singularity of finite coefficient coordinates.

The two models show why each step matters. For common-bath qubits in the zero- and single-excitation sector, every defective decaying mode contributes its full pole order to a suitable unconditional excitation--reference coherence. The EP3 example displays a quadratic Jordan term in one quadrature and its cancellation in another. For the fixed-photon-number optomechanical sector, the noise-history term and the Gaussian width combine two coefficient roots into a double-pole conditional displacement response at long preparation age and fixed lag on the regular normalizable branch. The purely algebraic coefficient tail cancels from the width, and complete averaging can turn conditional momentum narrowing into reduced-state broadening.

A coefficient EP therefore has a physical interpretation through a specified causal preparation, state reconstruction, and average. The exact NMQSD-to-NMME route makes those connections explicit without assigning a universal EP order to all observables or to the instantaneous reduced-state generator.

\begin{acknowledgments}
This material is based upon work supported by the National Science Foundation under Grant Nos. 2328948 and 2519355.
\end{acknowledgments}
\section*{Data availability}
The underlying data and code can be obtained from the corresponding author upon reasonable request.
\appendix
\section{Direct NMQSD closure and the non-Markovian master equation}
\label{q:app:prior}

In the retained zero- and single-excitation sector, take
$H_S=0\oplus H_1$ with time-independent Hermitian $H_1$, the
lowering operator in Eq.~\eqref{q:eq:L}, and the independent vacuum
reservoir specified in Sec.~\ref{q:sec:model}.
The operators in Eq.~\eqref{q:eq:qsd} act on the whole retained
system space, including their zero action on $\ket{G_S}$.
Products of two lowering operators vanish in this sector, so the
noise commutator in the consistency equation vanishes.
The remaining commutators give Eq.~\eqref{q:eq:f}, with boundary
condition $\bm f_q(t,t)=\kappa$.
Differentiating the integral in Eq.~\eqref{q:eq:qsd}, using
$\alpha(t,t)=A_0$ and $\partial_t\alpha(t,s)=-d\alpha(t,s)$,
then yields the Riccati equation \eqref{q:eq:riccati} and its
initial condition $\bm F_q(0)=0$.

Because $\bar O_t$ is noise independent, the exact projector
average in Eq.~\eqref{fw:nmme} gives Eq.~\eqref{q:eq:NMME}.
Direct multiplication of the ground--excitation blocks yields
Eq.~\eqref{q:eq:rho_blocks}. Integrating those equations with the
amplitude propagator in Eq.~\eqref{q:eq:FG} gives
\begin{align}
 \rho_{\mathrm{ee}}(t)&=\mathsf G(t)\rho_{\mathrm{ee}}(0)\mathsf G(t)^\dagger,\nonumber\\
 \rho_{\mathrm{e}G}(t)&=\mathsf G(t)\rho_{\mathrm{e}G}(0),\nonumber\\
 \rho_{GG}(t)&=\rho_{GG}(0)+
       \tr\!\left[(I-\mathsf G^\dagger\mathsf G)\rho_{\mathrm{ee}}(0)\right].
 \label{q:eq:appmap}
\end{align}
In particular, the preparation
$(\ket{G_S}+\ket{u_0})/\sqrt2$ gives
$\rho_{\mathrm{e}G}(t)=\mathsf G(t)u_0/2$.
This is an unconditional system coherence obtained directly from
the NMME.

The same result follows at the NMQSD-state level.
The excitation amplitude is the deterministic vector
$\mathsf G(t)u_0$, while noise enters the ground amplitude with
zero reference mean. On a regular chart the functional derivative
then gives
$\bm f_q(t,s)=\kappa\mathsf G(s)\mathsf G(t)^{-1}$.
Thus the coefficient closure, state propagation, and NMME refer to
the same preparation and reservoir, without an additional physical
evolution model.

\section{State-memory lift, global chart, and pole visibility}
\label{q:app:proofs}

Define the row $\bm W=\bm F_q\mathsf G$ on a regular chart.
Using Eqs.~\eqref{q:eq:riccati} and \eqref{q:eq:FG} cancels
both the $i\bm F_qH_1\mathsf G$ terms and the quadratic terms,
yielding the coupled equations in Eq.~\eqref{q:eq:state_memory}.
Since $\bm W(0)=0$, this auxiliary variable is exactly the
original memory convolution,
\begin{equation}
 \bm W(t)=A_0\int_0^t e^{-d(t-s)}
                         \kappa\mathsf G(s)\,ds .
 \label{q:eq:appmemory}
\end{equation}
It introduces no additional system state or physical preparation.
Eliminating it gives the causal amplitude equation
\begin{equation}
 \dot{\mathsf G}(t)=-iH_1\mathsf G(t)
 -\int_0^t\alpha(t,s)\kappa^\dagger\kappa\mathsf G(s)\,ds.
 \label{q:eq:appVolterra}
\end{equation}
The constant matrix $\mathsf A_q$ and initial block column
$(I_{N_q},0)^T$ are given in Eq.~\eqref{q:eq:state_memory}.
Thus $\mathsf G$ and $\bm W$ are blocks of the globally defined
matrix $U_q(t)=e^{\mathsf A_qt}$.

\subsection{Dual lifts and the finite coefficient chart}

Direct block multiplication establishes Eq.~\eqref{q:eq:similarity},
where $J^{-1}=J$. Its spectral map is $\mu=-\lambda-d$, with
$\mu$ and $\lambda$ eigenvalues of $M_q$ and $\mathsf A_q$,
respectively.
The transpose and invertible similarity preserve Jordan orders.
The propagators obey
$e^{M_qt}=e^{-dt}JU_q(t)^{-T}J$.
Because $\det U_q=e^{(-i\tr H_1-d)t}$, the complementary
minor formula gives
\begin{equation}
 q(t)=e^{-dt}(U_q^{-1})_{aa}
     =e^{i\tr(H_1)t}\det\mathsf G(t).
 \label{q:eq:appdet}
\end{equation}
This identity is global: $U_q$ is invertible at every
finite time, even when its system block is singular.
On $\det\mathsf G\ne0$, $\bm F_q=\bm W\mathsf G^{-1}$.
An isolated order-$m$ zero of $\det\mathsf G$ gives
$c_q=-\dot q/q\sim-m/(t-t_0)$.
It is a boundary of the finite coefficient chart.

\subsection{A positive metric and global reduced-state regularity}

The Hermiticity of $H_1$ and $A_0,\gamma>0$ give the positive metric
\begin{equation}
 \begin{gathered}
 P_q=\diag(I_{N_q},A_0^{-1}),\\
 \mathsf A_q^\dagger P_q+P_q\mathsf A_q
       =-\frac{2\gamma}{A_0}e_ae_a^\dagger .
 \end{gathered}
 \label{q:eq:appmetric}
\end{equation}
Applying this identity to the block column in
Eq.~\eqref{q:eq:state_memory} and integrating yields
\begin{equation}
 \begin{aligned}
 \mathsf G^\dagger\mathsf G&+\frac{\bm W^\dagger\bm W}{A_0}\\
 &+\frac{2\gamma}{A_0}\int_0^t
        \bm W(s)^\dagger\bm W(s)\,ds=I_{N_q}.
 \end{aligned}
 \label{q:eq:appcontraction}
\end{equation}
Hence $\mathsf G^\dagger\mathsf G\preceq I$.
The reduced map \eqref{q:eq:appmap} stays finite and completely
positive across a coefficient-chart zero.
For example, a Kraus representation consists of
$K_0=\ket{G_S}\bra{G_S}+\sum_{ij}\mathsf G_{ij}\ket i\bra j$
and ground-transfer rows obtained from a factorization of
$I-\mathsf G^\dagger\mathsf G$.
They satisfy $\sum_\alpha K_\alpha^\dagger K_\alpha=I$.
The time-local NMME uses finite coefficients only on its regular
chart; Eq.~\eqref{q:eq:appmap} supplies the continuous global map.

For an eigenvector $v=(u,w)^T$ of $\mathsf A_q$,
Eq.~\eqref{q:eq:appmetric} gives
\begin{equation}
 \operatorname{Re}\lambda=-\gamma
 \frac{|w|^2/A_0}{\|u\|^2+|w|^2/A_0}\le0 .
 \label{q:eq:appeigenmetric}
\end{equation}
If $\operatorname{Re}\lambda=0$, then $w=0$, $\kappa u=0$, and
$-iH_1u=\lambda u$.
Moreover $v^\dagger P_q$ is a left eigenvector at $\lambda$.
A generalized vector satisfying
$(\mathsf A_q-\lambda I)v_1=v$ would imply
$0=v^\dagger P_qv>0$, a contradiction.
These boundary modes are therefore dark and semisimple.
Every defective eigenvalue has $\operatorname{Re}\lambda<0$.

\subsection{Causal reachability and a surviving system pole}

For $\operatorname{Re}\lambda<0$, the matrix $\lambda I+iH_1$ is invertible.
The eigenvector equations fix
$u=-(\lambda I+iH_1)^{-1}\kappa^\dagger w$ uniquely, with
$w\ne0$. Each such eigenspace is one dimensional; an algebraic
multiplicity $r$ therefore corresponds to one size-$r$ block.
The Laplace transform of the prescribed coefficient denominator
is given by Eq.~\eqref{q:eq:qres}.
At $\mu=-\lambda-d$ its numerator equals
$(-1)^{N_q}\det(\lambda I+iH_1)\ne0$.
Thus the zero-coefficient initial condition and denominator readout
retain the full order of every defective pole.
The coefficient tail follows from Eq.~\eqref{fw:tail} only when
that reached pole is strictly dominant.
Writing $\Gamma_\lambda=-\operatorname{Re}\lambda$ gives
$\operatorname{Re}\mu=\Gamma_\lambda-\gamma$:
coefficient dominance selects faster amplitude decay.

For $D_q(z)$ and $h_q(z)$ defined in
Eq.~\eqref{q:eq:schur_functions}, the Schur complement of
$zI-\mathsf A_q$ gives the system resolvent block in
Eq.~\eqref{q:eq:schur}.
At a defective eigenvalue $\lambda$ of order $r$,
$\det D_q(\lambda)\ne0$ and $h_q$ has an order-$r$ zero.
Expanding $h_q^{-1}$ about that zero gives the highest system
Laurent coefficient in Eq.~\eqref{q:eq:residue}.
Both outer-product factors are nonzero: otherwise invertibility
would imply $\kappa=0$, which cannot produce a defective
eigenvalue in this class.
Consequently,
\begin{equation}
 \mathsf G(t)\supset
 \frac{t^{r-1}e^{\lambda t}}{(r-1)!}(R_r)_{\mathrm{ee}}.
 \label{q:eq:appvisible}
\end{equation}
Some normalized system input $u_0$ and system row $v^\dagger$
have $v^\dagger(R_r)_{\mathrm{ee}}u_0\ne0$.
The NMME coherence $\mathsf G(t)u_0/2$ then exposes the term in
a suitable Hermitian quadrature.
This argument also covers an EP embedded in a larger spectrum.
It does not assign the same order to every quadrature or to the
complete density-operator generator.

\section{Exact EP3 solution and coefficient spectral splitting}
\label{q:app:benchmark}

Use the two-qubit parameters of Eq.~\eqref{q:eq:params} and
$\tau=\gamma t$. Their relation to the configuration of
Ref.~\cite{lin2025} is explicit under the real orthogonal basis
rotation
\begin{equation*}
 \begin{gathered}
 U_{30}=\frac12\begin{pmatrix}\sqrt3&1\\-1&\sqrt3\end{pmatrix},
 \qquad \kappa U_{30}=(0,\sqrt{8/27}),\\
 U_{30}^T H_1U_{30}=\frac{\gamma}{3\sqrt3}
                   \begin{pmatrix}0&1\\1&0\end{pmatrix}.
 \end{gathered}
\end{equation*}
Thus the exchange magnitude is $\gamma/(3\sqrt3)$ and the total
bath coupling weight is $A_0\|\kappa\|^2=8\gamma^2/27$,
corresponding to $\Lambda=\gamma$ and $\Gamma=16\gamma/27$ in
that reference. The following calculation uses the original
NMQSD coefficient lift.
Direct evaluation gives
\begin{equation}
 \det(\mu I-M_q)=(\mu+2\gamma/3)^3.
 \label{q:eq:appchar}
\end{equation}
For $\mathcal N_M=M_q+2\gamma I_3/3$,
$\mathcal N_M^3=0$ and
$(\mathcal N_M^2)_{aa}=4\gamma^2/27\ne0$.
Hence its nilpotency index is three, establishing the EP3
directly in the coefficient equations.
Exponentiating this block gives Eqs.~\eqref{q:eq:qEP} and
\eqref{q:eq:cEP}. The positive coefficients of the polynomial in
Eq.~\eqref{q:eq:qEP} ensure a regular chart for $t\ge0$, and its
quadratic degree yields the stated $-2/t$ correction to $c_q$.

The derived state-memory matrix has
$\mathcal N_A=\mathsf A_q+\gamma I_3/3$ with
$\mathcal N_A^3=0$, so its system block is
\begin{equation}
 \mathsf G(t)=e^{-\tau/3}
 \left[I_3+\frac{\tau}{\gamma}\mathcal N_A+
              \frac{\tau^2}{2\gamma^2}\mathcal N_A^2\right]_{\mathrm{ee}}.
 \label{q:eq:appGexact}
\end{equation}
The diagonal similarity $R=\operatorname{diag}(1,\sqrt3,\sqrt6)$ makes $R^{-1}\mathsf A_qR$ rational-complex at $\gamma=1$. Exact arithmetic gives ranks $2,1,0$ for the first three powers of its shifted nilpotent matrix; the affine-transpose identity gives the coefficient result.

For $u_0=(1,0)^T$, its two components are
\begin{align}
 a_1(t)&=e^{-\tau/3}
          \left[1+\frac{(2-i)\tau}{6}-\frac{i\tau^2}{18}\right],
 \nonumber\\
 a_2(t)&=-\frac{\sqrt3}{54}e^{-\tau/3}
                              \tau[(2+i)\tau+3i].
 \label{q:eq:appamplitudes}
\end{align}
The coherent preparation gives $\rho_{jG}=a_j/2$.
Thus $X_1$ has a linear prefactor and $Y_1$ retains the quadratic
term, as in Eq.~\eqref{q:eq:quadratures}. The selected signal in
Eq.~\eqref{q:eq:X2pure} has Laplace transform
\begin{equation}
 \widetilde X_2(s)=-\frac{2\sqrt3\,\gamma^2}
                         {27(s+\gamma/3)^3}.
 \label{q:eq:appX2}
\end{equation}
The nonzero cubic pole is therefore a consequence of the direct
NMQSD-to-NMME map.
For the pure initial excitation $\ket1$, the total excitation population $P_{\rm exc}=\tr\rho_{\rm ee}$ is
\begin{equation}
 P_{\rm exc}(t)=e^{-2\tau/3}
 \frac{2\tau^4+6\tau^3+36\tau^2+162\tau+243}{243}.
 \label{q:eq:population}
\end{equation}
Its fourth-degree prefactor comes from products of amplitudes and
does not by itself identify a fifth-order EP.

Under the frequency shift defined above Eq.~\eqref{q:eq:cubic},
the coefficient characteristic polynomial in that equation gives
$\zeta^3\sim\varepsilon/9$.
The state-memory eigenvalues obey
$(\lambda+\gamma/3)/\gamma=-\zeta$ and have the same pairwise
spectral gaps.
This splitting is derived from $M_q$; the state map fixes its
readout consequences.
Numerical route comparisons and their finite-grid accuracy are
reported separately in Appendix~\ref{q:app:numerics}.

\section{Three coefficient and state boundary cases}
\label{q:app:controls}

All three controls use $A_0=\gamma^2$, $\Omega=0$, and
$\tau=\gamma t$.
They follow directly from the same coefficient and NMME
equations.

\emph{Dark degeneracy.}
Take $N_q=3$, $H_1=\gamma\diag(0,0,1)$, and
$\kappa=(0,0,1/2)$.
The first two excitation states have constant amplitudes.
Their repeated eigenvalue $\lambda=0$ of $\mathsf A_q$ is
semisimple, as is the corresponding $\mu=-\gamma$ of $M_q$.
The coefficient eigenvectors have last coordinate zero and
therefore lie outside the finite stationary-row chart.
This does not mean that the prescribed trajectory $q(t)$
vanishes identically.

\emph{Off-EP chart pole.}
Take $N_q=1$, $H_1=0$, and $\kappa=1$.
The two eigenvalues of $M_q$ are the distinct values
$\mu_\pm=-\gamma/2\pm i\sqrt3\gamma/2$.
With $\beta_0=\sqrt3/2$,
\begin{align}
 q(t)=\mathsf G(t)&=e^{-\tau/2}
  \left(\cos\beta_0\tau+\frac{\sin\beta_0\tau}{\sqrt3}\right),
 \nonumber\\
 \dot q(t)&=-\gamma e^{-\tau/2}
                            \frac{\sin\beta_0\tau}{\beta_0}.
 \label{q:eq:appchartcontrol}
\end{align}
The first zero is $t_0=4\pi/(3\sqrt3\,\gamma)$, with
$\dot q(t_0)=-\gamma e^{-\gamma t_0/2}\ne0$.
Thus $c_q\sim-1/(t-t_0)$ without a coefficient EP.
Equation~\eqref{q:eq:appmap} gives the regular ground-state
density at $t_0$ for both an initial excitation and its equal
superposition with $\ket{G_S}$.
The global map continues through this point; no finite
coefficient value is assigned there.

\emph{Reached but subleading EP.}
Take
$H_1=\gamma\diag(1/(4\sqrt2),-1/(4\sqrt2))$ and
$\kappa=(3/8,3/8)$.
\begin{align}
 q(t)&=\frac92e^{-\tau/2}
       -\left(\frac72+\frac{3\tau}{8}\right)e^{-3\tau/4}
       \nonumber\\
       &=e^{-\tau/2}Q(\tau),\nonumber\\
 Q(\tau)&=\frac92-\left(\frac72+\frac{3\tau}{8}\right)
                                                e^{-\tau/4},\nonumber\\
 Q'(\tau)&=\left(\frac12+\frac{3\tau}{32}\right)
                                                e^{-\tau/4}>0.
 \label{q:eq:appsubleading}
\end{align}
Here $q(t)$ is the prescribed denominator.
The coefficient spectrum has a size-two block at
$\mu=-3\gamma/4$ and a simple pole at $\mu=-\gamma/2$.
Since $Q(0)=1$, the chart is finite for all $t\ge0$.
The EP contribution is reached, but
$c_q=\gamma[1/2-Q'(\tau)/Q(\tau)]\to\gamma/2$
has no leading $1/t$ term.
The state-memory eigenvalues are the defective
$\lambda=-\gamma/4$ and simple $\lambda=-\gamma/2$.
Thus the faster-decaying amplitude mode selects the coefficient
tail, while the EP polynomial remains explicitly present in
the global state propagation.

\section{Noise-complete two-time coefficients}
\label{o:app:O}
The oscillator terms below follow the noise-complete $O$-operator equations of Ref.~\cite{dgs1998}, with the conserved affine term included. The integrated coefficients $F,G,R$ are defined in Sec.~\ref{o:sec:model}; the scalar-history definitions are
\begin{align}
\mathcal J(t,u)&=\int_0^t\alpha(t,s)j(t,s,u)\dd s,\nonumber\\
\Xi_t&=\int_0^t\mathcal J(t,u)z_u^*\dd u,\nonumber\\
J(t)&=\int_0^t\alpha(t,u)\mathcal J(t,u)\dd u.
\label{o:eq:historydefinitions}
\end{align}
Matching coefficients in Eq.~\eqref{o:eq:consistency} gives
\begin{align}
\partial_tf&=\nu g_O-2\ii\ell Fg_O+\ii\ell Gf-\ii\ell\mathcal J(t,s),\nonumber\\
\partial_tg_O&=-\nu f-\ii\ell Gg_O,\nonumber\\
\partial_tr_O&=\eta g_O-\ii\ell Rg_O,\nonumber\\
\partial_tj(t,s,v)&=-\ii\ell g_O(t,s)\mathcal J(t,v).
\label{o:eq:two_time}
\end{align}
The endpoint conditions are
\begin{align}
f(t,t)&=\ell,&g_O(t,t)&=r_O(t,t)=0,\nonumber\\
j(t,t,v)&=0,&j(t,s,t)&=\ell g_O(t,s).
\label{o:eq:boundaries}
\end{align}
The last condition comes from the upper limit of the noise integral and $[Lz_t^*,g_OP]$. All integrated coefficients initially vanish. Since $\partial_t\alpha(t,s)=-d\alpha(t,s)$ for $t>s$,
\begin{equation}
\partial_t\mathcal J(t,v)=-(d+\ii\ell G)\mathcal J(t,v),\qquad
\mathcal J(t,t)=\ell G(t).
\label{o:eq:Jhistory}
\end{equation}
Differentiating the integrals in Eq.~\eqref{o:eq:historydefinitions} gives $\dot J=A_0\ell G-(2d+\ii\ell G)J$ and the $\Xi$ equation in Eq.~\eqref{o:eq:Gaussian}. Integration of the first line of Eq.~\eqref{o:eq:two_time} combines $-2\ii\ell FG+\ii\ell GF$ into $-\ii\ell FG$, yielding Eq.~\eqref{o:eq:coeff}. These identities hold on regular coefficient charts; a ratio-coordinate singularity is continued through the homogeneous variables.

\section{Coefficient exceptional point and Gaussian response}
\label{o:app:spectral}
\label{o:app:width}
Shifting the spectral variable in Eq.~\eqref{o:eq:Pa} gives
\begin{equation}
P_a(w-d)=(w^2-d^2)(w^2+\nu^2)+2\ii B\nu d.
\label{o:eq:shiftedpoly}
\end{equation}
A multiple root requires $\Delta_a=0$ or $w=0$. The latter would require $d\nu=2\ii B$, impossible for $\gamma>0$. Write $y=\Omega^2$ and $s_a=\nu^2+\gamma^2-y$. The imaginary and real parts of $\Delta_a=0$ give $B=\Omega s_a/(2\nu)$ and $s_a^2+4y(\nu^2-y)=0$. Their only positive solution is Eq.~\eqref{o:eq:locus}, for which $\nu^2<y<\nu^2+\gamma^2$.

The geometric multiplicities can be read from a finite-matrix similarity. Set $s_0=\sqrt{2A_0}$, $g_a=\sqrt{2B}$, and
\begin{equation}
S=\begin{pmatrix}0&1&0&0\\-1&0&0&0\\
0&0&-\ii/s_0&A_0/s_0\\0&0&-1/s_0&\ii A_0/s_0
\end{pmatrix}.
\end{equation}
Direct multiplication gives $\det S=1$ and
\begin{equation}
S({M_o}+d\Id)S^{-1}
=\begin{pmatrix}0&\nu&0&0\\-\nu&0&-g_a&0\\
0&0&0&-\ii d\\-g_a&0&\ii d&0\end{pmatrix}
\equiv K_{\rm alg}.
\label{o:eq:similarity}
\end{equation}
Each eigenspace of $K_{\rm alg}$ is one-dimensional: its first coordinate fixes the others when $\nu,g_a,d\ne0$. Hence the two repeated roots $\mu_\pm=-d\pm w$ are defective. This similarity is used only to establish the Jordan structure of the coefficient matrix.

Causal reachability follows from the denominator generated by $v(0)=e_4$. Its Laplace transform is
\begin{equation}
\widetilde q(s)=\frac{\mathcal N_o(s)}{P_a(s)},\qquad
\mathcal N_o(s)=(s+2d)[(s+d)^2+\nu^2]-\ii B\nu.
\label{o:eq:qLaplace}
\end{equation}
At any root $\mu$ of $P_a$, $\mathcal N_o(\mu)=-\ii B\nu(\mu+2d)/\mu\ne0$. Thus
\begin{equation}
q(t)=(\mathcal C_{+}t+\mathcal C_{0})e^{\mu_+t}+(\mathcal C_{-}t+\mathcal C_{1})e^{\mu_-t},
\qquad \mathcal C_{+}\ne0.
\label{o:eq:qJordan}
\end{equation}
Using $G=\dot q/(\ii\ell q)$ gives Eq.~\eqref{o:eq:Gtail}. Equivalently, the initial vector is cyclic because $\det[e_4,M_o e_4,{M_o}^2e_4,{M_o}^3e_4]=-(A_0\ell)^4\nu^2\ne0$. The constants in Eq.~\eqref{o:eq:qJordan} are coefficients of the homogeneous solution and are unrelated to the wavefunction scalar $c_{\psi,t}$.

\paragraph*{Width cancellation and the response kernel.}
For the root $w=x_w+\ii y_w$ with $x_w>0$, the identities $2x_wy_w=\gamma\Omega$ and $x_w^2-y_w^2=(\gamma^2-\Omega^2-\nu^2)/2$ give $y_w>0$ and
\begin{equation*}
\left(y_w^2-\frac{\Omega^2}{2}\right)
\left(1+\frac{\gamma^2}{2y_w^2}\right)=\frac{\nu^2}{2}>0.
\end{equation*}
Consequently, $2y_w>\Omega$ and $\Rea a_*=(2y_w-\Omega)/\nu>0$ for all positive $\gamma,\Omega,\nu$. The working interval $0<\gamma<\nu$ in Eq.~\eqref{o:eq:width} is therefore separate from stationary-width positivity and is used together with the regular normalizable branch hypothesis.

The coefficient equations give $F=\ii(\ddot q+d\dot q)/(\nu\ell q)$. At the leading EP root $\mu=w-d$, substitution of $a_*=\ii(d-2w)/\nu$ into the width equation yields
\begin{equation}
\left.\dot a\right|_{a=a_*}
=\frac{2\ii}{\nu}\frac{(\partial_t-\mu)^2q}{q}.
\label{o:eq:widthforcing}
\end{equation}
The differential operator annihilates the entire leading block $(\mathcal C_{+}t+\mathcal C_{0})e^{\mu t}$. Its forcing comes only from the exponentially subleading block. Of the two stationary widths, $a_*$ and $a_u=\ii d/\nu$, the latter has $\Rea a_u=-\Omega/\nu<0$.

For the relaxation estimate, put $a=\dot u/(\ii\nu u)$ and $u=qv$. With $\vartheta=t+\mathcal C_{0}/\mathcal C_{+}$,
\begin{equation}
\ddot v+\left[-w^2+\frac{2w}{\vartheta}-\frac{2}{\vartheta^2}
+O(e^{-2\Rea w t})\right]v=0.
\label{o:eq:widthv}
\end{equation}
The leading solutions are
\begin{align}
v_1&=\frac{e^{wt}}{\vartheta},\nonumber\\
v_2&=e^{-wt}\left(\frac{\vartheta}{2w}+\frac{1}{2w^2}
+\frac{1}{4w^3\vartheta}\right).
\label{o:eq:widthbasis}
\end{align}
The two leading solutions have a nonzero constant Wronskian. Variation of constants multiplies the remainder in Eq.~\eqref{o:eq:widthv} by $v_1^2$, $v_1v_2$, or $v_2^2$, with respective integrable late-time bounds $O(t^{-2})$, $O(e^{-2\Rea w\,t})$, and $O(t^2e^{-4\Rea w\,t})$. A vanishing $v_1$ amplitude would select $a_u$. For the normalized initial state, however, the exact NMQSD identity $\mathbb E_0\norm{\psi_t}^2=1$ requires the nonzero Gaussian kets to remain normalizable on the regular branch. Since $a$ is independent of the noise labels, this enforces $\Rea a>0$ and excludes convergence to $\Rea a_u<0$. The homogeneous coefficient variables continue across any earlier zero of $q$; Gaussian normalizability remains a separate condition. The surviving $v_1$ contribution gives Eq.~\eqref{o:eq:width}, including the polynomial--exponential correction.

At finite preparation age, the response of $(b,\Xi)^T$ is governed by
\begin{equation}
\begin{aligned}
A_b(t)&=\begin{pmatrix}\ii\ell G-\ii\nu a&-\ii\ell\\0&-d-\ii\ell G\end{pmatrix},\\
B_b(t)&=\binom{\ell}{\ell G(t)}.
\end{aligned}
\label{o:eq:finite_response}
\end{equation}
If $\Phi(t,s)$ is its fundamental propagator, the finite-age kernel is $h_T(s)=(1,0)\Phi(T,s)B_b(s)$. Equivalently, the row $v_T(s)$ satisfies $v_T'(s)=-v_T(s)A_b(s)$ with $v_T(T)=(1,0)$. The stationary-lag kernel in Eq.~\eqref{o:eq:nearEPkernel} is obtained only after the long-age limit, rather than by replacing $T$ with the lag.

For distinct $w_1,w_2$, the inverse transform of Eq.~\eqref{o:eq:transfer} is
\begin{equation}
\frac{h(\theta)}{\ell}
=\frac{(d-w_1)e^{-w_1\theta}-(d-w_2)e^{-w_2\theta}}{w_2-w_1}.
\label{o:eq:partialfractions}
\end{equation}
Combining the two terms gives Eq.~\eqref{o:eq:nearEPkernel}. Locally, $\bar w$ and $\epsilon_w^2$ are analytic in $\Delta_a$, while $\cosh(\epsilon_w\theta)$ and $\sinhc(\epsilon_w\theta)$ are even in $\epsilon_w$. Hence the kernel is regular at every fixed finite lag. At the EP, $A_\infty=-w\Id+\mathcal N_J$, with $(\mathcal N_J)_{12}=-\ii\ell$ and $\mathcal N_J^2=0$, which yields Eq.~\eqref{o:eq:EPkernel}. The coefficient $d-w$ cannot vanish: $w=d$ in $w^2=(d^2-\nu^2)/2$ would require $d^2=-\nu^2$, inconsistent with $\gamma,\Omega>0$. The same substitution in Eq.~\eqref{o:eq:shiftedpoly} excludes a canceled pole away from the EP when $B\nu d\ne0$.

\section{Paired Novikov average}
\label{o:app:average}
The coupled noise average and its adjoint follow the exact quantum-Brownian-motion construction of Ref.~\cite{strunz2004}. The correlation convention in Eq.~\eqref{o:eq:noise} fixes both contractions:
\begin{align}
\mathbb E_0[z_u\Pi_t]&=\int_0^t\alpha(u,v)\mathbb E_0[O(t,v)\Pi_t]\dd v,\nonumber\\
\mathbb E_0[z_u^*\Pi_t]&=\int_0^t\alpha(v,u)\mathbb E_0[\Pi_tO^\dagger(t,v)]\dd v.
\label{o:eq:Novikov}
\end{align}
At fixed endpoint $t$, set $A_{t,s}=f(t,s)Q+g_O(t,s)P+r_O(t,s)N$ and $Y_s=\mathbb E_0[O(t,s,z^*)\Pi_t]$. Novikov averaging couples $Y_s$ to its adjoint:
\begin{align}
Y_s&=A_{t,s}\rho_t+\int_0^t\mathcal K_t(s,v)Y_v^\dagger\dd v,\nonumber\\
\mathcal K_t(s,v)&=\ii\int_0^t j(t,s,u)\alpha(v,u)\dd u.
\label{o:eq:pairedkernel}
\end{align}
The adjoint equation is
\begin{equation}
Y_s^\dagger=\rho_tA_{t,s}^\dagger
+\int_0^t\overline{\mathcal K_t(s,v)}Y_v\dd v.
\label{o:eq:pairedadjoint}
\end{equation}
For $X_k\in\{Q,P,N\}$, expand $Y_s=\sum_k[u_k(s)X_k\rho_t+v_k(s)\rho_tX_k]$ and denote the corresponding coefficients of $A_{t,s}$ by $a_k(s)$. Since $Q,P,N$ are Hermitian, matching their left and right actions on $\rho_t$ gives
\begin{equation}
u_k=a_k+\mathcal K_tv_k^*,\qquad
v_k=\mathcal K_tu_k^*.
\label{o:eq:uvpair}
\end{equation}
Here $a_Q=f$, $a_P=g_O$, and $a_N=r_O$. The product $\mathcal K_t\overline{\mathcal K_t}$ means integral-operator composition with pointwise complex conjugation, without exchanging the kernel arguments; it is not $\mathcal K_t\mathcal K_t^\dagger$. Eliminating $v_k$ gives
\begin{equation}
u_k=(I-\mathcal K_t\overline{\mathcal K_t})^{-1}a_k,\qquad
v_k=\mathcal K_tu_k^*.
\label{o:eq:pairedinverse}
\end{equation}
Invertibility of $I-\mathcal K_t\overline{\mathcal K_t}$ specifies the time-local averaging chart; $\norm{\mathcal K_t}<1$ is sufficient, but is not assumed as a universal bound. This inverse supplies the present kernel representation of that paired closure.

Define
\begin{equation}
U_k=\int_0^t\alpha(t,s)u_k(s)\dd s,\qquad
V_k=\int_0^t\alpha(t,s)v_k(s)\dd s.
\label{o:eq:UV}
\end{equation}
Then $\mathcal A_t=\sum_k(U_kX_k\rho_t+V_k\rho_tX_k)$. Substitution into Eq.~\eqref{o:eq:formalME}, using $L=L^\dagger=\ell Q$, combines the coefficients into $U_k-V_k^*$ and gives Eqs.~\eqref{o:eq:closedME}--\eqref{o:eq:ABD}. The scalar-history correlations remain in these effective coefficients.

\section{Complete Gaussian norm integral}
\label{o:app:Gaussian}
For the finite Gaussian source in Appendix~\ref{o:app:numerics}, the linear displacement and quadratic scalar equations in Eq.~\eqref{o:eq:Gaussian} give
\begin{equation}
\begin{aligned}
b&=b_0+\beta^T\mathbf{x},\\
c_\psi&=c_{\psi,0}+\zeta^T\mathbf{x}+\frac12\mathbf{x}^T\Gamma\mathbf{x},\qquad \Gamma^T=\Gamma.
\end{aligned}
\label{o:eq:GaussianSource}
\end{equation}
Here $\mathbf{x}\sim N(0,\Id)$ is a finite real Gaussian source vector, distinct from the position coordinate in the wavefunction. The source-independent terms $b_0,c_{\psi,0}$ and response coefficients $\beta,\zeta,\Gamma$ depend on time. The Gaussian integrals retain the full scalar $c_\psi$. The squared norm of the stochastic ket, which supplies its statistical weight, is
\begin{equation}
\begin{aligned}
w(\mathbf{x})&=\norm{\psi_t}^2\\
&=\sqrt{\frac{\pi}{a_R}}
\exp\left[2\Rea c_\psi+\frac{(\Rea b)^2}{a_R}\right].
\end{aligned}
\label{o:eq:norm}
\end{equation}
Combining this weight with the unit Gaussian source measure gives the real precision matrix and linear term
\begin{align}
\mathsf P&=\Id-2\Rea\Gamma-\frac{2}{a_R}\beta_R\beta_R^T,\nonumber\\
q_x&=2\Rea\zeta+\frac{2\Rea b_0}{a_R}\beta_R.
\label{o:eq:precision}
\end{align}
Here $\beta_R=\Rea\beta$ and $\beta_I=\Ima\beta$ are real column vectors. When $\mathsf P\succ0$, completion of the square gives
\begin{equation}
\bar{\mathbf{x}}=\mathsf P^{-1}q_x,\qquad \Sigma=\mathsf P^{-1},
\label{o:eq:weightedsource}
\end{equation}
with absolute total weight
\begin{equation}
\begin{aligned}
Z&=\sqrt{\frac{\pi}{a_R}}\frac{\exp(\chi)}{\sqrt{\det\mathsf P}},\\
\chi&=2\Rea c_{\psi,0}+\frac{(\Rea b_0)^2}{a_R}
+\frac12q_x^T\mathsf P^{-1}q_x.
\end{aligned}
\label{o:eq:Z}
\end{equation}
The matrix mapping source fluctuations to the state centroid is
\begin{equation}
\mathsf B=T_a\begin{pmatrix}\beta_R^T\\\beta_I^T\end{pmatrix},\qquad
m_0=T_a\binom{\Rea b_0}{\Ima b_0}.
\label{o:eq:Bsource}
\end{equation}
On the positive-precision chart $a_R>0$, $\mathsf P\succ0$, the normalized finite-representation moments are
\begin{equation}
m=m_0+\mathsf B\bar{\mathbf{x}},\qquad
V=V_\psi+\mathsf B\Sigma\mathsf B^T.
\label{o:eq:GaussianMoments}
\end{equation}
At a finite source resolution $M_{\rm src}$, these moments use
\begin{equation}
\dd\mathbb P_t^{(M_{\rm src})}
=\frac{w_{M_{\rm src}}}{Z_{M_{\rm src}}}\,\dd\mathbb P_0.
\label{o:eq:finiteweight}
\end{equation}
The continuous exact model has $Z=1$. At finite resolution, the absolute weight $Z_{M_{\rm src}}$ is checked separately from the normalization used to compute the moments. Retaining $c_{\psi,0}$ and its time derivative is essential for the absolute checks of $\log Z$ and $\partial_t\log Z$. Analytic time differentiation of these Gaussian expressions supplies the moment derivatives used below.

\section{Qubit readout selection and direct numerical comparisons}
\label{q:app:numerics}

\subsection{Direct NMQSD and NMME calculations}
\label{q:app:direct}

\begin{figure*}[tbp]
\centering
\includegraphics[width=\textwidth]{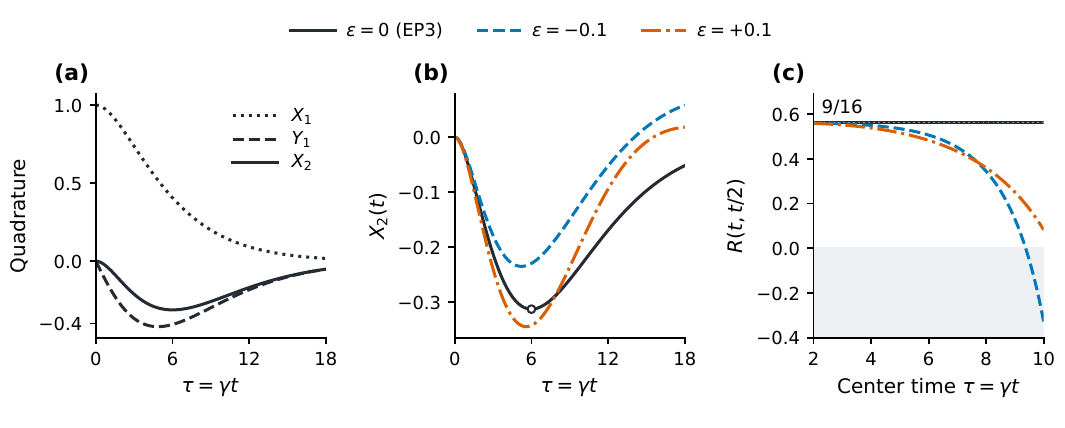}
\caption{Local quadrature selection within the exact NMME. The preparation is $(\ket{G_S}+\ket1)/\sqrt2$, with the benchmark parameters of Eq.~\eqref{q:eq:params} and a fixed rotating-phase reference. (a) At the EP, $X_1$ has a linear polynomial prefactor, $Y_1$ has linear and quadratic terms, and $X_2$ selects the quadratic term. (b) Direct NMME evolution of $X_2$ for the EP and the two shifts $\varepsilon=\epsilon/\gamma=\pm0.1$. The circle marks the EP maximum magnitude at $\tau=6$. (c) The signed three-time ratio at the same $161$ centers for each evolution. The EP value is $9/16$; the detuned controls depart at finite times. For $R\leq0$, the real logarithmic degree $\ln R/\ln(3/4)$ is undefined. This ratio is a model-dependent diagnostic, not a general EP criterion.}
\label{q:fig:local_selection_ratio}
\end{figure*}

We propagate system amplitudes and the full system density operator. We set $\gamma=A_0=1$, $\Omega=0$, and use the three shifts $\epsilon/\gamma=0,-0.1,+0.1$ of Eq.~\eqref{q:eq:params}. Two initial states, $(\ket{G_S}+\ket j)/\sqrt2$ with $j=1,2$, determine both columns of the system propagator. The nonlinear coefficient row in Sec.~\ref{sec:qubits} is integrated first; its dense interpolant drives separate integrations of $\dot{\mathsf G}=(-iH_1-\kappa^\dagger\bm F_q)\mathsf G$ and of the full $3\times3$ system density operator in Eq.~\eqref{q:eq:NMME}. The two routes share the exact coefficient closure and propagate different objects. Their agreement tests the internal NMQSD-to-NMME calculation. Figure~\ref{q:fig:local_selection_ratio} shows selected coherence channels and their finite-time ratio.

The integrations use DOP853 \cite{hairer1993} with relative tolerance $2\times10^{-12}$ and absolute tolerance $2\times10^{-14}$. We save $7201$ uniformly spaced times over $0\leq\tau=\gamma t\leq18$, with spacing $0.0025$. Across both preparations, all system-coherence components, and the three shifts, the maximum sampled absolute difference between NMQSD and NMME is $8.50\times10^{-12}$. At the EP, the maximum absolute discrepancy of the doubled quadratures $X_1,Y_1,X_2$ from their analytic expressions is $1.28\times10^{-12}$. All reported numerical discrepancies and conditioning diagnostics are discrete samples on the stated grids; they provide neither continuous-time error bounds nor a guarantee of invertibility between samples.

The system block of $e^{\mathsf A_qt}$, derived from the internal coordinate $\bm W=\bm F_q\mathsf G$, agrees with the NMQSD amplitude matrix within $1.69\times10^{-11}$ and with the NMME coherence reconstruction within $6.44\times10^{-13}$ on the same grid. Density-matrix trace and Hermiticity discrepancies remain below $7\times10^{-16}$, at roundoff scale. The minimum sampled density eigenvalue is $-7.93\times10^{-13}$, at the numerical error scale; complete positivity of the exact map follows from the analytic contraction argument in Appendix~\ref{q:app:proofs}.

A separate coefficient-only integration reaches $\tau=1000$ without dividing underflowing homogeneous coordinates. Its maximum sampled absolute discrepancy in $c_q/\gamma$ is $2.07\times10^{-12}$, and the final value of $t[2\gamma/3-\operatorname{Re}c_q(t)]$ is approximately $1.99$.

\subsection{Selected quadrature and finite-time ratio}

The selected $X_2$ signal in Eq.~\eqref{q:eq:X2pure} gives a
one-channel diagnostic. For a single component
$S(t)=At^m e^{-\beta t}$,
\begin{equation}
 \begin{aligned}
 R(t,h)&=\frac{S(t-h)S(t+h)}{S(t)^2}\\
       &=\left[1-(h/t)^2\right]^m,\qquad t>h>0.
 \end{aligned}
 \label{q:eq:ratio}
\end{equation}
The denominator must be nonzero. A constant gain and the exponential factor cancel. For the EP signal $X_2(t)=-(\sqrt3/27)\tau^2e^{-\tau/3}$, this gives
\begin{equation}
 R(t,t/2)=\frac{X_2(t/2)X_2(3t/2)}{X_2(t)^2}
 =\frac{9}{16}.
 \label{q:eq:numerical_ratio}
\end{equation}
Figure~\ref{q:fig:local_selection_ratio} compares this signed ratio from the unconditional NMME quadratures at $161$ centers $2\leq\tau\leq10$ for the EP and both detunings. For the negative and positive shifts, respectively, its values at $\tau=6$ are approximately $0.51$ and $0.48$; at $\tau=10$ they are approximately $-0.33$ and $0.08$.

For these noiseless, model-dependent readouts, additive offsets, a shifted time origin, phase mixing, and a small denominator can alter the ratio. Both detuned controls also begin quadratically, which precludes using the short-time ratio alone as an EP test. Reconstructing the full coefficient row requires the separately prepared ensembles in Eq.~\eqref{q:eq:coherenceG}.

\subsection{Coefficient reconstruction and conditioning}
\label{q:app:inverse}

Equations~\eqref{q:eq:inverseF} and \eqref{q:eq:inversec}
reconstruct the specified coefficients from the separate
preparations and phase-calibrated coherences in
Eq.~\eqref{q:eq:coherenceG}, on an invertible chart.

A derivative-free interval form follows on a path where the
determinant never vanishes:
\begin{align}
 \frac1{\Delta t}\int_t^{t+\Delta t}c_q(u)\,du
 &=-i\tr H_1\nonumber\\
 &\quad-\frac1{\Delta t}\operatorname{Log}_{\rm cont}
    \frac{\det\mathsf G(t+\Delta t)}{\det\mathsf G(t)}.
 \label{q:eq:interval}
\end{align}
The logarithm follows the determinant phase continuously;
nonzero endpoints alone do not establish a regular path.
Pointwise differentiation amplifies measurement and discretization
errors, while small singular values of $\mathsf G$ amplify the
inverse. A condition number alone is not an absolute guard:
a scalar map has condition number one at every nonzero value.
The reconstructed contraction is compared with its coefficient
solution using the direct integrations in Appendix~\ref{q:app:direct}.

\begin{figure}[htbp]
\centering
\includegraphics[width=\columnwidth]{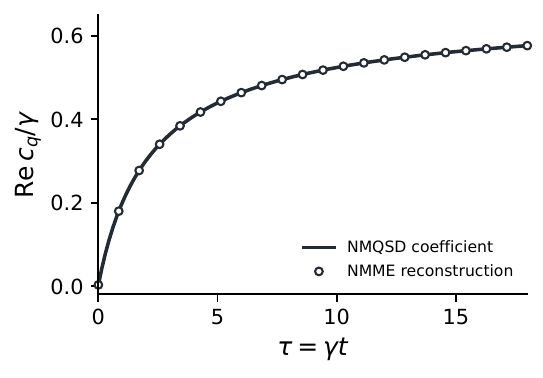}
\caption{System readouts recover the coefficient contraction at the EP. The line gives the direct NMQSD coefficient solution $\operatorname{Re}c_q/\gamma$. Open circles reconstruct the contraction from the exact NMME coherences for the two preparations $(\ket{G_S}+\ket j)/\sqrt2$, $j=1,2$, using a five-point derivative. Both calculations use the same specified $H_1$, coupling convention, and phase reference. The plot is noiseless; the inverse is model dependent and requires an invertible system propagator.}
\label{q:fig:coherence_inverse}
\end{figure}

The numerical reconstruction uses $\mathsf G_{ij}=2\rho_{iG}^{(j)}$ from each NMME preparation. At the $7193$ admissible centers $0.01\leq\tau\leq17.99$, a centered five-point derivative with step $h=0.005/\gamma$ supplies $\dot{\mathsf G}$ in Eqs.~\eqref{q:eq:inverseF} and \eqref{q:eq:inversec}. Figure~\ref{q:fig:coherence_inverse} and Table~\ref{q:tab:direct_inverse} compare the resulting $\bm F_q$ and $c_q$ with the separately integrated coefficient solution.

\begin{table}[htbp]
\caption{Sampled errors in the noiseless NMME-based coefficient reconstruction. In the first row, the maximum is over the $7193$ admissible times and both components of $\Delta\bm F_q$; the second row uses the same times. The third row takes the maximum over all $7201$ saved times on $0\leq\tau\leq18$, including the endpoints.}
\label{q:tab:direct_inverse}
\centering
\begin{ruledtabular}
\begin{tabular}{lrrr}
$\epsilon/\gamma$&$0$&$-0.1$&$+0.1$\\
\hline
$\max|\Delta\bm F_q|/\gamma$&$1.26\times10^{-10}$&$3.17\times10^{-9}$&$3.80\times10^{-11}$\\
$\max|\Delta c_q|/\gamma$&$1.85\times10^{-11}$&$1.63\times10^{-9}$&$4.15\times10^{-11}$\\
$\max\operatorname{cond}_2\mathsf G$&$64.3$&$752$&$7.31$\\
\end{tabular}
\end{ruledtabular}
\end{table}

The minimum sampled singular values of $\mathsf G$ over the full time grid are approximately $1.9\times10^{-3}$, $4.6\times10^{-4}$, and $1.1\times10^{-2}$ for the EP, negative shift, and positive shift. The exact EP denominator in Eq.~\eqref{q:eq:qEP} is nonzero for all $t\geq0$.

\FloatBarrier
\section{Finite-time oscillator coefficient and state comparisons}
\label{o:app:validationfigures}
\begin{figure*}[tbp]
\centering
\includegraphics[width=\textwidth]{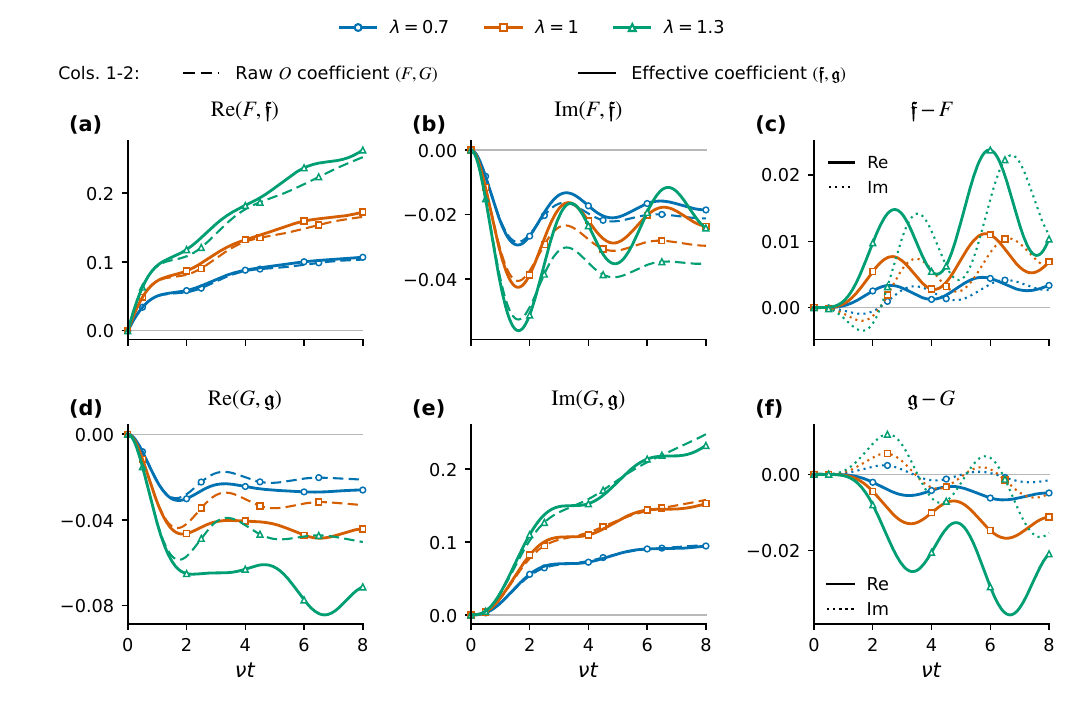}
\caption{The noise average changes the coefficients that govern the reduced state. The upper and lower rows show $F,\mathfrak f$ and $G,\mathfrak g$, respectively, for $\lambda=0.7,1,1.3$. (a),(b),(d),(e) real and imaginary parts of the raw $O$-operator coefficients (dashed) and effective master-equation coefficients (solid); (c),(f) their noise-induced corrections, with real parts solid and imaginary parts dotted. Parameter identities are also distinguished by circles, squares, and triangles. The imaginary parts of $\mathfrak f$ and $\mathfrak g$ enter the restoring force and damping, while their real parts enter diffusion. Parameters are those of Fig.~\ref{o:fig:branches}, with $a(0)=1$, $b(0)=0$, and the normalized vacuum wavefunction. The three fixed couplings compare finite-time dynamics across the coefficient exceptional point.}
\label{o:fig:coefficients}
\end{figure*}

\begin{figure*}[tbp]
\centering
\includegraphics[width=\textwidth]{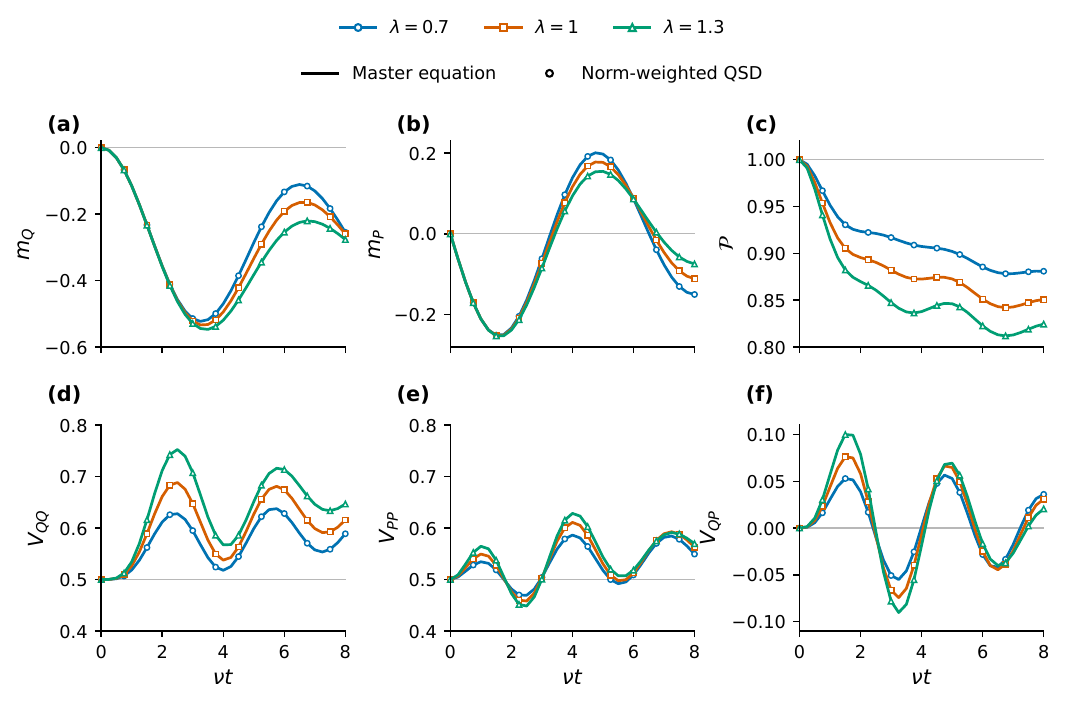}
\caption{Reduced-state evolution from the same Gaussian initial state. Solid lines show independent integration of the master-equation moment equations; hollow symbols show the norm-weighted NMQSD average at every third saved time and the final endpoint. The panels give the two means, purity $\mathcal P=1/(2\sqrt{\det V})$, and three independent covariance entries. Their agreement tests the complete noise average over $0\leq\nu t\leq8$ for the three fixed couplings. The shared initial state has $m(0)=0$ and $V(0)=\Id_2/2$; other parameters are those of Fig.~\ref{o:fig:branches}.}
\label{o:fig:dynamics}
\end{figure*}

Figures~\ref{o:fig:coefficients} and \ref{o:fig:dynamics} compare the coefficients and full Gaussian-state trajectories for the preparation in Sec.~\ref{o:sec:results}; Table~\ref{o:tab:finite_state} gives their endpoint values.

The complete average changes the coefficients that act on the density operator. Figure~\ref{o:fig:coefficients} shows the original $F,G$, the effective $\mathfrak f,\mathfrak g$, and the corrections separately. For example, at $\nu t=8$ and $\lambda=1$, $\Rea G\approx-0.033$ whereas $\Rea\mathfrak g\approx-0.044$. This change enters the cross diffusion in Eq.~\eqref{o:eq:ABD}. The corresponding imaginary parts modify the deterministic motion.

The full-state trajectories in Fig.~\ref{o:fig:dynamics} show oscillatory mean motion while the initially pure reduced state becomes mixed. The three couplings produce different displacement, covariance, and purity trajectories over the common interval. For a single-mode Gaussian state, the purity is \mbox{$\mathcal P=\tr\rho^2=1/(2\sqrt{\det V})$} \cite{adesso2014}. Across these three couplings at $\nu t=8$, stronger coupling lowers the purity and raises the position variance. These finite-time comparisons do not by themselves establish an exceptional point of the reduced-state generator.
\begin{table}[htbp]
\caption{Reduced Gaussian state at $\nu t=8$. The three fixed couplings share the initial state and bath frequency specified in Sec.~\ref{o:sec:results}. Values are rounded from the 256-mode norm-weighted NMQSD averages, with refinement checks in Appendix~\ref{o:app:numerics}.}
\label{o:tab:finite_state}
\setlength{\tabcolsep}{4pt}
\begin{ruledtabular}\begin{tabular}{cccccc}
$\lambda$ & $m_Q$ & $m_P$ & $V_{QQ}$ & $V_{PP}$ & $\mathcal P$\\
0.7 & -0.255 & -0.150 & 0.589 & 0.549 & 0.881\\
1.0 & -0.260 & -0.111 & 0.616 & 0.563 & 0.851\\
1.3 & -0.277 & -0.075 & 0.647 & 0.570 & 0.824\\
\end{tabular}\end{ruledtabular}
\end{table}

\section{Numerical representation and convergence}
\label{o:app:numerics}
The diagnostics below concern the stated preparation and finite comparison grids. Both propagation routes share the exact coefficient core, so their agreement tests the subsequent averaging and moment propagation; it is not an absolute error bound for the continuous exact solution. The source-kernel discrepancy is likewise evaluated on the specified grid.

On the fixed interval $0\leq t\leq T$, with $T=8/\nu$, the source uses $M_{\rm src}=128,256$ filtered temporal modes and an independent initial mode. The corresponding dimensions are 129 and 257 complex coordinates, or 258 and 514 real coordinates, respectively. The temporal basis consists of the orthonormal shifted Legendre functions
\begin{equation}
e_j(t)=\sqrt{\frac{2j+1}{T}}\,P_j(2t/T-1).
\label{o:eq:sourcebasis}
\end{equation}
Here $j=0,\ldots,M_{\rm src}-1$ and $P_j$ is the Legendre polynomial of degree $j$. Their filtered components satisfy
\begin{equation}
\dot\phi_j(t)=e_j(t)-d^*\phi_j(t),\qquad \phi_j(0)=0.
\label{o:eq:filteredsource}
\end{equation}
The finite colored source is
\begin{equation}
\begin{aligned}
z_{M_{\rm src}}^*(t)={}&\sqrt{A_0}e^{-d^*t}\xi_0\\
&+\sqrt{2\gamma A_0}\sum_{j=0}^{M_{\rm src}-1}\phi_j(t)\xi_{j+1},
\end{aligned}
\label{o:eq:finitecoloredsource}
\end{equation}
where the $\xi_k$ are independent standard proper complex Gaussian variables, with $\mathbb E_0[\xi_k\xi_l^*]=\delta_{kl}$ and \mbox{$\mathbb E_0[\xi_k\xi_l]=0$}. The independent initial term retains the stationary Ornstein--Uhlenbeck initial distribution. The real source in Eq.~\eqref{o:eq:GaussianSource} is obtained by writing $\xi_k=(x_{2k+1}+\ii x_{2k+2})/\sqrt{2}$, with independent $x_j\sim N(0,1)$. The norm integral retains $c_\psi(t,z^*)$ and its derivative and is evaluated analytically, without random draws.

Writing $z_{M_{\rm src}}^*(t)=\sum_k s_k(t)\xi_k$, the reconstructed source covariance is
\begin{equation}
\alpha_{M_{\rm src}}(t,s)=\sum_k s_k^*(t)s_k(s).
\label{o:eq:sourcecovariance}
\end{equation}
In the numerical units $\nu=1$, the kernel comparison grid $\mathcal T$ is the union of the 33 output times $\{0,0.25,\ldots,8\}$ and the 128-point Gauss--Legendre nodes on $[0,8]$. Its maximum-entry discrepancy is
\begin{equation}
\epsilon_{\rm ker}^{(M_{\rm src})}
=\frac{\max_{t,s\in\mathcal T}|\alpha_{M_{\rm src}}(t,s)-\alpha(t,s)|}
{\max\{1,\max_{t,s\in\mathcal T}|\alpha(t,s)|\}}.
\label{o:eq:kernelmetric}
\end{equation}
For $M_{\rm src}=256$, this diagnostic ranges from approximately $4.13\times10^{-4}$ to $7.68\times10^{-4}$ across the three couplings.

The master-equation calculation starts from the same initial state and the same exact coefficient core as the NMQSD calculation, then independently computes the paired average and propagates the moments. The paired average uses order-64 quadrature at 129 coefficient times, and order-128 comparisons at $\nu t=0.5,1,2,4,6,8$. The largest recorded condition number of the discretized paired-averaging matrix $I-\mathcal K_t\overline{\mathcal K_t}$ in Eq.~\eqref{o:eq:pairedinverse} is below $1.6$.

The master-equation propagation compares coefficient spacings $\nu\Delta t=0.125$ and $0.0625$, with relative and absolute integration tolerances $10^{-10}$ and $10^{-12}$. Real and imaginary cubic interpolation supplies the coefficients between their saved times. The 33 common output times are separated by $\nu\Delta t=0.25$. Interpolation is used for propagation; the figures connect the original saved coefficient or state nodes.

For a compared quantity $X$, define
\begin{equation}
\epsilon(X)=\frac{\norm{X-X_{\rm ref}}_{\max}}
{\max(1,\norm{X_{\rm ref}}_{\max})}.
\label{o:eq:error}
\end{equation}
The reference is the finer representation for source and mesh comparisons, and the 256-mode norm-weighted NMQSD result for the master-equation comparison. The denominators are one for the reported moment discrepancies; these are not percentages relative to each small component. Table~\ref{o:tab:convergence} separates source refinement, time-mesh refinement, and agreement of the two propagation routes.

The absolute norm checks at the higher source resolution give $|\log Z|<10^{-7}$ and $|\partial_t\log Z|<10^{-7}$. Over the 33 common output times for each coupling, the scaled discrepancies between the finer master-equation trajectory and the 256-mode NMQSD average are below $10^{-7}$ in both $m$ and $V$. The instantaneous covariance-derivative discrepancy is also below $10^{-7}$. The largest 128/256 source difference among $m,V,\dot m,\dot V$ is below $10^{-6}$.

Physicality checks cover the reduced, conditional, and centroid covariances. In the convention $[Q,P]=\ii$, the Gaussian quantum condition \cite{adesso2014} is $V+\ii J_2/2\succeq0$, where $J_2=\left(\begin{smallmatrix}0&1\\-1&0\end{smallmatrix}\right)$, and $V_{\rm centroid}\succeq0$. The saved Gaussian states satisfy these conditions within numerical tolerance; the master-equation output and accepted adaptive states are checked as well. Purity uses valid covariance nodes, including every displayed node. These sampled-state checks do not prove continuous-time physicality or complete positivity for all preparations.

\begin{table}[htbp]
\caption{Maximum covariance discrepancies over the 33 output times. The columns compare source modes 128/256, master-equation coefficient spacings $\nu\Delta t=0.125/0.0625$, and the finer master-equation trajectory against the 256-mode NMQSD average, respectively. Discrepancies use Eq.~\eqref{o:eq:error}.}
\label{o:tab:convergence}
\begin{ruledtabular}\begin{tabular}{cccc}
$\lambda$ & $\epsilon_{\rm src}(V)$ & $\epsilon_{\rm mesh}(V)$ & $\epsilon_{\rm ME-QSD}(V)$\\
0.7 & $1.59\times10^{-7}$ & $4.34\times10^{-7}$ & $3.05\times10^{-8}$\\
1.0 & $2.67\times10^{-7}$ & $6.26\times10^{-7}$ & $4.62\times10^{-8}$\\
1.3 & $4.16\times10^{-7}$ & $8.24\times10^{-7}$ & $6.00\times10^{-8}$\\
\end{tabular}\end{ruledtabular}
\end{table}

\end{document}